\documentclass[aps,prl,twocolumn,preprintnumbers,
superscriptaddress,floatfix]{revtex4}

\usepackage{amssymb,multirow,amsmath}
\usepackage{graphicx,color}

\begin{document}

\newcommand{\Tr}{\mbox{Tr\,}}
\newcommand{\beq}{\begin{equation}}
\newcommand{\eeq}{\end{equation}}
\newcommand{\bea}{\begin{eqnarray}}
\newcommand{\eea}{\end{eqnarray}}
\renewcommand{\Re}{\mbox{Re}\,}
\renewcommand{\Im}{\mbox{Im}\,}

\title{Domain Wall Fermions on the Brane}

\author{Jes\'us Cruz Rojas}
\affiliation{ STAG Research Centre \&  Physics and Astronomy, University of
Southampton, Southampton, SO17 1BJ, UK}

\author{Nick Evans}
\affiliation{ STAG Research Centre \&  Physics and Astronomy, University of
Southampton, Southampton, SO17 1BJ, UK}

\author{Jack Mitchell}
\affiliation{ STAG Research Centre \&  Physics and Astronomy, University of
Southampton, Southampton, SO17 1BJ, UK}

\begin{abstract}
We study  domain wall fermions and their condensation in the D3/probe D7 system.   A spatially dependent mass term for the ${\cal N}=2$ hypermultiplet can be arranged to isolate distinct two component fermions on two 2+1 dimensional domain walls. We argue that the system shows condensation/mass generation analogous to the D3/probe D5 $\overline{\rm D5}$ system. The chiral condensate and pion mass can be directly computed on the domain wall. We provide evidence that these systems with the domains separated by a width $w$ have a bare (current) quark mass that scales as $1/w$ when the spatial dependent mass is large. Adding a magnetic field does not induce chiral symmetry breaking between the separated domain wall fermions, but a similar phenomenological dilaton factor can be made strong enough to introduce spontaneous symmetry breaking.  We show a Gell-Man-Oakes-Renner relation for the pions in that case and also for the case where the D7 probe is in a back-reacted dilaton flow geometry. The vacuum configurations can also be interpreted as  having a spontaneously generated mass by a Nambu-Jona-Lasinio four fermion operator, depending on the choice of boundary conditions on fluctuations, according to Witten's multi-trace prescription. 
\end{abstract}

\maketitle

\section{I Introduction} \vspace{-0.5cm}

Kaplan introduced the idea of domain wall fermions in \cite{Kaplan:1992bt} to address the challenging issue of including chiral fermions in lattice gauge theory simulations. A spatially dependent mass term for quarks with a sharp crossing from positive to negative mass isolates a single chirality of a Dirac spinor at the boundary. The technique has gone on to become an important tool for the lattice community.

In this paper we will apply the ideas of domain wall fermions but in holography \cite{Maldacena:1997re}. Here we are motivated by enlarging the holographic toolbox of methods to construct particular gauge theories with holographic descriptions, as well as the formal aspect of studying a new environment in holography. 

As a first step in this direction we will consider spatially dependent masses in the D3/probe D7 system which is a very well understood holographic construction \cite{Karch:2002sh,myers}. The base D3/probe D7 system describes an ${\cal N}=2$ quark hypermultiplet in the fundamental representation interacting with the adjoint fields of ${\cal N}=4$ super Yang Mills theory. Here domain walls, on which the quark mass is zero, isolate distinct two component sub-spinors of the four dimensional theory at each domain wall. In 2+1 dimensions there is no formal chirality projector but the system is very analogous to the usual reduction performed - we will briefly review this formalism in Section III. Note that some holographic work on a single  defect of this type already exists in \cite{HoyosBadajoz:2010ac} in a condensed matter setting. Here though we will study the interactions between defects. 

In the holographic setting, we begin by studying (co)sinusoidal quark mass terms in a direction $x_3$ or $z$ in the field theory. To introduce the ideas we begin by working at the level of the linearized equations of motion for fluctuations of the D7 brane to allow analytic computation without complicated partial differential equations. The results in this approximation contain many of  the physics ideas we would expect. Each Fourier mode of momentum $k$ is associated with a wave function $f_k$ in the holographic radial direction on the probe, $\rho$. The higher $k$, the quicker these modes die off into the infra-red, i.e. small $\rho$. We can then construct arbitrary $z$ dependent functions for the quark mass, $M$, as Fourier series. We choose a periodic function with two close domain walls, well separated from the next pair. Since the higher Fourier modes decouple as one moves to small $\rho$ the two domain walls, defined by where $M=0$  approach each other and eventually join. The locus where $M=0$ in the $z-\rho$ plane is very reminiscent of well known U-shaped brane anti-brane systems with chiral symmetry breaking such as the Sakai-Sugimoto model \cite{Sakai:2004cn} or the D3/probe D5 $\overline{D5}$ system \cite{Karch:2000gx,Skenderis:2002vf}. Thus we interpret the domain wall system as showing symmetry breaking - for a single probe D7 there is a U(1) symmetry on the spinors on each domain wall that is broken to the ``vector" sub-group. In the large spatial-dependent mass limit, the presence of a single two component fermion on the domain wall effectively removes the coupling to the adjoint scalar of the ${\cal N}=4$ theory - if there were many branes then one would expect a full U($N_f$) symmetry to be present on each domain wall and the IR joining of the walls to break these symmetries to a single U($N_f)$ group.  Here it remains unclear as to whether the breaking is explicit or spontaneous.

Having introduced the set up in the linearized approximation, we then move to the opposite limit where the 3+1 dimension mass, $M$, is infinite except on the domain wall junction. We show how to solve for the position of the domain wall in the $z-\rho$ plane and again recover U-shaped configurations. Interestingly, the equation for these configurations precisely reproduces the action of the D3/probe D5 $\overline{D5}$ system presumably showing that the symmetry breaking is identical between the two systems. 

The quark anti-quark operator/source pair that control the symmetry breaking, when seen from a UV perspective, are Wilson lines with a quark on each end (one on each domain wall)\cite{Aharony:2008an}. These are stringy states and as in brane models not directly accessible. However, in the IR where the string becomes of zero length these operators should be indistinguishable from the 3+1d theory's quark condensate with which they will mix. One would expect the D7 brane embedding solution to display a quark condensate as the normalizable piece of the UV solution, localized on the brane. We restrict the equation of motion for the 3+1d probe scalar, that describes this operator/source pair, to the domain wall locus and then solve.  It provides a direct local measure of the quark condensate and its phase should play the role of the Goldstone of the symmetry breaking. 

For the case of a D7 probe in pure AdS$_5$, the domain wall configurations have a single associated scale, the width separating the two defects, $w$. We show it sets all scales including the minimum $\rho$ to which the $M=0$ contour reaches and the ``Goldstone'' mass. Here the Goldstone is not massless and so we conclude that this system has a bare (or current) quark mass of $1/w$. Given the similarities to the D3/probe D5 $\overline{D5}$ system it is likely that system does too. 

To further investigate chiral symmetry breaking in the system we apply a baryon number magnetic field. This is easily introduced in the D7 probe system via a world-volume gauge field and enters as an effective dilaton factor multiplying the action. Magnetic fields are known to induce dynamical mass generation \cite{Gusynin:1995nb,Filev:2007gb} although whether the dynamics is strong enough to overcome the spatial separation of the domain walls on which the participating fermions reside requires computation. The linearized analysis breaks down in this system because the magnetic field induces chiral symmetry breaking in the full 3+1 dimensions for the light quarks \cite{Filev:2007gb}. In the large $M$ limit though only the domain walls, where $M=0$, feel the IR presence of the magnetic field. We find that the contour where $M=0$ is pushed to higher scales, but it still remains the case that there are configurations of large width where the minimum $\rho$ of the U-shaped domain wall configuration approaches zero. The magnetic field renormalizes the bare mass but does not induce spontaneous symmetry breaking. Again the U-shapes we find are precisely those of the D3/probe D5 $\overline{D5}$ system in a magnetic field and so that system is expected to also not show dynamical symmetry breaking. 

Inspired by the magnetic field case we try phenomenological choices of a multiplying dilaton factor in the action that blow up more strongly as $\rho \rightarrow 0$. Here, for suitable choices, we show that there is a minimum value of $\rho$ to which the U-shaped configurations reach even as the width between the domain walls diverges. This signals chiral symmetry breaking. Turning to the solutions of the holographic fields on the domain wall we also find that chiral symmetry breaking is now triggered consistently. We show a Gell-Mann-Oakes-Renner relation \cite{GellMann:1968rz} with the Goldstone mass proportional to the square root of the UV quark mass.

To show the dynamics in a more rigorous setting, we investigate the domain walls on a D7 brane in a dilaton flow geometry \cite{Constable:1999ch}.  The geometry is a solution of the supergravity equations of motion with a dilaton growing into the IR. The geometry has an IR pole and it is unclear whether it can be sensibly resolved in string theory, but it does provides a backreacted hard wall in the geometry. Such geometries are known to induce chiral symmetry breaking for the D7 probe theory in 3+1 dimensions \cite{Babington:2003vm,Ghoroku:2004sp} and that dynamics is, at least, away from the singularity. Again in the large $M$ limit the domain wall set up only sees the IR geometry along the $M=0$ defect so only the lower dimensional theory is affected. We again show that the U-shaped domain wall configurations reach down to only a minimum value of $\rho$ (above the singularity scale) and that the holographic fields on the domain wall consistently describe chiral symmetry breaking and a Gell-Mann-Oakes-Renner relation. 

We will also discuss Witten's multi-trace prescription \cite{Witten:2001ua} (see \cite{Evans:2016yas} for its application in the D3/ probe D7 system) which implies that any vacuum configuration with a source present can be reinterpreted as a system with no source but a higher dimension operator causing condensation. In the case of a quark mass one imagines a Nambu-Jona-Lasinio (NJL) four fermion term \cite{Nambu:1961tp} is present causing a quark condensate. The combination of the condensate and four fermion operator  then dynamically generates a boundary quark mass term. In this case one must adjust the boundary conditions on fluctuations since the quark mass itself must fluctuate with the quark condensate. In the domain wall models under these conditions the Goldstone indeed becomes massless reflecting the dynamical symmetry breaking by the NJL term. 

Finally we will briefly overview 1+1 dimensional domain walls that can be generated in the D3/probe D5 system. The 1+1 dimensional fermions are expected to be chiral. Here, in the large mass limit, the equation for the domain wall locus matches that of a D3 probe in AdS$_5$. The behaviour of this system  is in other respects the same as the D3/probe D7 domain walls we have discussed in detail. 

The paper is organised as: in section II we will briefly overview U-shaped configurations in the D3/probe D5 $\overline{D5}$ system that we will use as a comparator system through the paper. In section III we review the domain wall fermion mechanism. In section IV we will enact domain walls in the D3/probe D7 system, first at the level of the linearized equations of motion and then in the large mass limit; we will discuss the Goldstone boson's nature and mass relations. In section V we include a magnetic field and dilaton profiles and show the conditions necessary for dynamical symmetry breaking. In Section VI we discuss Witten's multi-trace prescription in this setting. In Section VII we briefly summarize the D3/probe D5 system and 1+1 dimensional domain wall systems. We summarize and conclude in section VIII.

\section{II D3/probe D5 $\mathbf{\overline{D5}}$ comparator system}

We will be investigating a domain wall theory that consists of two 2+1d domain walls with massless fermionic fields that condense in the IR due to gauge interactions in the bulk.  We can realize a very similar system using a D3/ probe D5 $\overline{D5}$ system \cite{Karch:2000gx,Skenderis:2002vf}. We will make comparisons to this system in our analysis so we will review it first. 

The gravity dual of ${\cal N}=4$ SYM theory, the theory on the surface of a stack of $N_c$ D3 branes, is described by the near horizon AdS$_5\times$ S$^5$ geometry \cite{Maldacena:1997re}
\begin{equation}  \label{ads}
ds^2=\frac{r^2}{R^2}dx^{2}_{3+1}+\frac{R^2}{r^2}(d\rho^2+\rho^2d\Omega_2^2+du_1^2+du_2^2+ du_3^2)
\end{equation} 
where $R$ is the AdS radius and $r^2=\rho^2 + \sum_i u_i^2$. 

The probe D5-branes are arranged as
\begin{center}
 \begin{tabular}{c|c c c c c c c c c c}
      &0&1&2&3&4&5&6&7&8&9   \\   \hline
      D3&-&-&-&-& $\bullet$ & $\bullet$ &$\bullet$ &$\bullet$ &$\bullet$ &$\bullet$ \\
      D5/$\overline{\rm D5}$ &-&-&-&$\bullet$&-&-&-& $\bullet$& $\bullet$ &$\bullet$ 
 \end{tabular}
\end{center}\vspace{-1.8cm}

\begin{equation} \end{equation} \hspace{0.1cm}

with the D5 and $\overline{\rm D5}$ separated in the 3 direction ($z$).
The matter content on each of the 2+1d defects are two 2-component fermions plus scalar super partners interacting with the bulk ${\cal N}=4$ gauge theory. Interactions occur between the fermions on the domain walls leading the D5 and $\overline{\rm D5}$ to join \cite{Skenderis:2002vf}. The action is
\begin{equation} S_{D5} = -T_5 \int d^6\xi \sqrt{-det P[G_{MN} + 2 \pi \alpha' F_{MN}]}\end{equation}
which gives for $F_{MN}=0$
\begin{equation} S_{D5}\approx \int d\rho \,\, \rho^2\sqrt{1+\rho^4 z'^2} \label{d5act}\end{equation}
Note we have rescaled $x_{3+1}$ by a factor of $R^2$ to effectively set $R^2=1$. This means all momenta and masses below are strictly $R^4 k^2, R^4 M^2$. There is a conserved quantity so
\begin{equation} \frac{\rho^6 z'}{\sqrt{1+\rho^4 z'^2}} = \rho_{min}^4 \label{fff} \end{equation}
with U-shaped configuration solution
 \begin{equation}  \label{d5sol} z(\rho)= \pm Re\left[\frac{i}{\rho}\,\,_2 F_1\left[-\frac{1}{8},\frac{1}{2},\frac{7}{8},\left(\frac{\rho}{\rho_{min}}\right)^8\right]\right]
\end{equation}
The minimum $\rho$ value the U-shape reaches to, $\rho_{\rm min}$, determines the IR mass gap. The theory has only the width of the U-shape, $w$, as a UV parameter and the mass gap is given by
\begin{equation} r_{\rm min} = {0.675 \over w} \end{equation}

We can also switch on a B field \cite{Evans:2013jma} (absorbing factors of $2 \pi \alpha'$ and $R$) in the D3/probe D5 system in either of the $2,3$ spatial directions.
The action is then 
\begin{equation} S_{D5}\approx \int d\rho \,\, \sqrt{1 + {B^2  \over (\rho^2 + L^2)^2}} \rho^2\sqrt{1+\rho^4 z'^2}\end{equation}
The magnetic field prefactor  pushes $\rho_{\rm min}$ of the U-shaped configuations to higher values for a given width but $\rho_{\rm min}$ still approaches zero as the width of the U increases to infinity.  

We will use this system as a comparator for domain wall theories, returning to it in plots and discussion below.

\section{III Domain Wall Fermions}

In this section we briefly review the theory of domain wall fermions \cite{Kaplan:1992bt} concentrating on the particular case of a 3+1 dimensional theory dimensionally reduced to 2+1 dimensions.  Note the domain wall approach is usually used to reduce an odd dimensional theory (eg 4+1d) to an even dimensional theory (eg 3+1d) where the domain wall fermions are chiral. In our case where the domain wall is of odd dimension (emerging from an even dimension theory) there is no chirality projector but the higher dimension Dirac spinor is still split into two pieces each of which can be isolated on a separate brane. We will be happy with this as an example of the principle.

Consider the Dirac equation for a  3+1 dimensional free fermion with a mass that depends on $x_3$
\begin{equation} \left[ -i \gamma^\mu \partial_\mu -i \gamma^3 \partial_3 + M(x_3) \right] \Psi = 0 \label{4dDE} \end{equation}
where $\mu=0,1,2$. 
Under dimensional reduction $\Psi$ will become two 2+1 dimensional 2-component spinors which can be extracted from $\Psi$ by using the projectors
\begin{equation} P_{\pm} = \frac{1}{2} ( 1 \pm i \gamma_3) \end{equation}

Note in the Dirac basis 
\begin{equation} \begin{array}{cc} \gamma^0 = \left( \begin{array}{cccc}  1 & 0 & 0 & 0 \\ 0 & 1 & 0 & 0 \\ 0 & 0 & -1 & 0 \\ 0 & 0 & 0 & -1 \end{array}  \right),   
\gamma^1 = \left( \begin{array}{cccc}  0 & 0 & 0 & 1 \\ 0 & 0 & 1 & 0 \\ 0 &  -1 & 0 & 0 \\ -1 & 0 & 0 & 0 \end{array}  \right),  \\ & \\
\gamma^2 = \left( \begin{array}{cccc}  0 & 0 & 0 & -i \\ 0 & 0 & i & 0 \\ 0 & i & 0 & 0 \\ -i & 0 & 0 & 0 \end{array}  \right),   
\gamma^3 = \left( \begin{array}{cccc}  0 & 0 & 1 & 0 \\ 0 & 0 & 0 & -1 \\ -1 &  0 & 0 & 0 \\ 0 & 1 & 0 & 0 \end{array}  \right)  \end{array}
\end{equation}
If we write ($\pm$ correspond to the static energy eigenvalue)
\begin{equation}  \Psi = \left( \begin{array}{c} \psi^+_1 \\ \psi^+_2 \\ \psi^-_1 \\ \psi^-_2 \end{array} \right) \end{equation}
then the two distinct spinors each with the information of a 2-spinor are
\begin{equation}  P_\pm \Psi = \frac{1}{2}
\left( \begin{array}{c} \psi^+_1 \pm i \psi^-_1 \\ \psi^+_2 \mp i  \psi^-_2 \\ -i(\psi^+_1 \pm i \psi^-_1) \\ i(\psi_2^+ \mp i \psi^-_2) \end{array} \right)
\end{equation}

Now consider the case where 
\begin{equation} 
M(x_3) = -M, ~~   x_3<0, \hspace{1cm}  M(x_3) = M, ~~   x_3>0
\end{equation}
To seek a massless mode solution of (\ref{4dDE}), we decompose $\Psi$ in terms of a product in the $x^3$ and $x^\mu$ directions. 
\begin{equation}
\Psi =  \left[a(x_3) P_+ + b(x_3) P_- \right] \psi_0(x^\mu)  \end{equation}
where we assume  the massless eigenstate satisfies
\begin{equation}  \label{3dDE} i \gamma^\mu \partial_\mu \psi_0(x^\mu) = 0 \end{equation}

Since $\{ \gamma^\mu, \gamma^3\} =0$ we have $\gamma^\mu P^+ = P^- \gamma^\mu$ and $\gamma^\mu P^- = P^+ \gamma^\mu$ and we may drop the first term in (\ref{4dDE}) as a result of  (\ref{3dDE}) for the zero mode. 

Now we use $(\gamma^3)^2=-1$ so that  $i\gamma^3 P_+ = P_+$ and $i \gamma^3 P_- = - P_- $. The coefficients of $P_\pm$ give the two equations
\begin{equation} (\partial_3  + M(x_3) )a(x_3) = 0 \hspace{1cm}  (-\partial_3  + M(x_3)) b(x_3) = 0 \end{equation}
The first equation (remember $M(x_3)$ switches sign at the origin) has the normalizable  solution 
\begin{equation} a(x_3) = N e^{-M |x_3|}  \end{equation}
The solution for $b$ which has a positive sign in the exponential is not normalizable so unphysical. Thus a single one of the two 2+1 dimensional 2-spinors is massless at the discontinuity.  If we have a second discontinuity with the opposite sign switch in $M(x_3)$ then the second 2+1d spinor will be localized there. 

Note that a condensate between the two 2-spinors $\bar{\psi_1} \psi_2$ with the 2+1d $\gamma_0=\sigma_3$ is the same combination of operators as the 3+1d condensate $\bar{\Psi} \Psi$.

At weak coupling there is expected to be a quark mass controlled by the overlap of the wave functions so it will fall off as an exponential of the gap between two adjacent discontinuities (formally as exp$(-M w)$ with $w$ the separation between the defects). It is not clear that the same decoupling will happen if the separated quarks are interacting strongly - indeed we will find the mass in the holographic setting falls off only as the power law $\sim 1/w$. Our goal for the rest of the paper is to realize this domain wall set up in holography at strong coupling in part to investigate such questions.

\section{IV The D3/probe-D7 system \& Domain walls} 

For this section we rewrite the metric of the  gravity dual of ${\cal N}=4$ SYM theory as
\begin{equation}  \label{ads}
ds^2=\frac{r^2}{R^2}dx^{2}_{3+1}+\frac{R^2}{r^2}(d\rho^2+\rho^2d\Omega_3^2+du_1^2+du_2^2)
\end{equation} 
where $R$ is the AdS radius and $r^2=\rho^2 + \sum_i u_i^2$.

We introduce a probe ${\cal N}$=2 quark hypermultiplet into the ${\cal N}=4$ SYM theory described by (\ref{ads}) by including a D7 brane in the configuration \cite{Karch:2002sh}
\begin{center}
 \begin{tabular}{c|c c c c c c c c c c}
      &0&1&2&3&4&5&6&7&8&9   \\   \hline
      D3&-&-&-&-& $\bullet$ & $\bullet$ &$\bullet$ &$\bullet$ &$\bullet$ &$\bullet$ \\
      D7&-&-&-&-&-&-&-&-&$\bullet$ &$\bullet$ 
 \end{tabular}
\end{center}
\vspace{-1.7cm}

\begin{equation} \end{equation} \vspace{0.1cm}

The Dirac-Born-Infeld (DBI) action for the probe D7 is given by
\begin{equation} S_{D7} = -T_7\int d^8\xi \sqrt{-det P[G_{MN}]} \end{equation}
which gives up to constants
\begin{equation} S_{D7} \approx \int d^4x ~d\rho\,\, \rho^3 \sqrt{1+(\partial_\rho u_i)^2+\frac{R^4}{(\rho^2 + u_i^2)^2}(\partial_x u_i)^2} \label{dupe}
\end{equation}
where $u_i, i=1,2$ are the position of the brane in the two transverse directions. The rotational symmetry allows us to assume the vacuum embedding lies in the $u_1$ direction.  $x$ are generically the 3+1d Minkowski coordinates.
Note it is helpful again to rescale $x_{3+1}$ by a factor of $R^2$ to
effectively set $R^2=1$. This means all momenta and masses below are strictly $R^4 k^2, R^4 M^2$. 

\subsection{Mass terms in the linearized system} 

A supersymmetry preserving, $x$ independent mass can be included by the solution $u_1=m$ with $m_q=m/2 \pi \alpha'$. The meson spectrum for this case has been computed in \cite{myers}.

Here we will be interested in $z$ (i.e. $x_3$) dependent mass terms. To allow in principle a generic $z$ dependence we will write the mass term as a Fourier Series in terms of sine and cosine waves in the $z$ direction. Initially we will apply the mass as a perturbation to the massless theory, working in the linearized equation of motion approximation (keeping only terms to quadratic order in the action). That is we have
\begin{equation} \label{linu1}
\partial_\rho\left(\rho^3\partial_\rho u_1\right)+\frac{1}{\rho}(\partial_{z}^2 u_1)=0
\end{equation}
and we seek solutions 
\begin{equation}
u_1 = f_k(\rho) \cos kz
\end{equation}
Since there is no scale in the AdS geometry the solutions for $f_k(\rho)$ are ill behaved in the IR. We resolve this by including a hardwall regulator at $\rho=1$ - we shoot from $\rho=1$ with $u_1'(1)=0$. In practice this means that any structure we see in AdS can only be trusted for $\rho \gg 1$.

The numerical solutions asymptotes to a constant value in the UV (the UV form of the solution is $u_1 \sim m + c/\rho^2$). 
We then normalize the solutions so that 
\begin{equation}f_k(\rho \rightarrow \infty) = 1\end{equation}
Note that in the linearized regime the solutions are independent of the normalization and hence the physics is independent of the maximum mass value. 
 We plot some example $f_k$ in Figure 1. We see that higher $k$ modes are less supported at small $\rho$ as one would expect since UV physics is irrelevant in the IR.  \vspace{-0.5cm}

 \begin{center}
    \includegraphics[width=9cm]{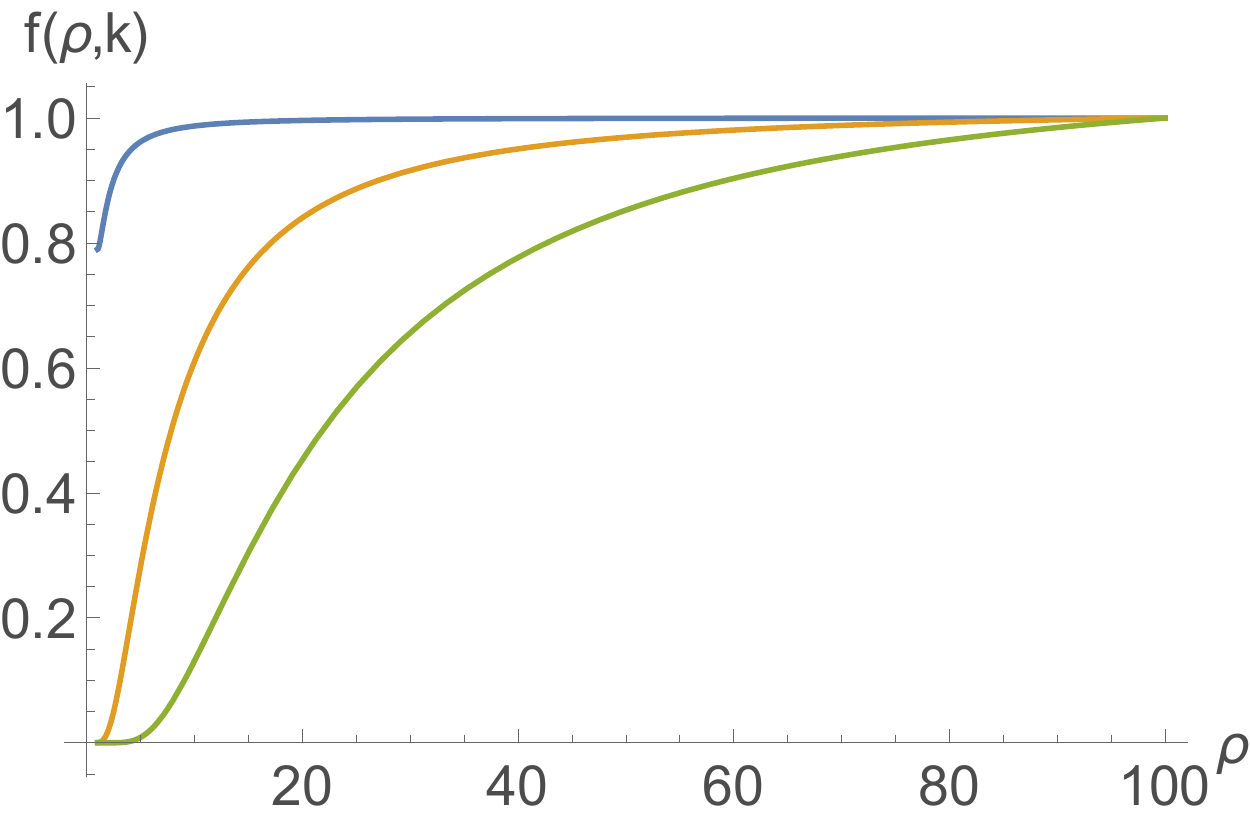} 
 \vspace{-0.4cm}
 
\noindent{{\textit{Figure 1: The solutions for $f_k(\rho)$ in pure AdS with a hardwall at $\rho=1$ for $k=1 ({\rm blue}),10 ({\rm orange}),30({\rm green})$.  }}}
\end{center} \vspace{-0.2cm}

\subsection{A Periodic Domain Wall Configuration} \vspace{-0.3cm} 

We now want to construct a configuration of two interacting domain walls. For simplicity we will use the following periodic example which is simple to Fourier expand. The 
 configuration is  3L periodic, with a sharp wall at $z_0$ and another at $3L-z_0$. The defect is centred half way along the $3L$ interval and is of width $w=3L-2 z_0$. Thus
\begin{equation} \begin{array}{lc}  u_1 = 1  \hspace{1cm}& 0 \leq z \leq z_0 \\ & \\ u_1= -1   & z \leq z \leq 3L-z_0 \\ & \\u_1 = 1 & 3 L - z_0   \leq z \leq 3L \end{array}\end{equation}
For $z_0 >L$ the walls are reasonably close but well separated from the next recurrence of the configuration - we will not take $z_0>L$ therefore in what follows.

The Fourier expansion for this even function is 
\begin{equation} \begin{array}{ccl}
    f(z)& =& \frac{a_0}{2}+ \sum_{n=1}^{\infty} a_n \cos{\frac{2\pi nz}{3L}}\\ &&\\
    a_0 &=& \frac{8z_0}{3L}-2\\ &&\\
    a_n &=& \frac{2}{\pi n}   \bigg[ \sin \frac{2\pi n z_0}{3L}- \sin 
    \frac{2\pi n(3L-z_0)}{3L}  \bigg]\\ 
\end{array} \end{equation}

In Figure 2 we plot an example configuration showing the Fourier approximation  taking the first 100 terms.

\begin{center}
    \includegraphics[width=9cm]{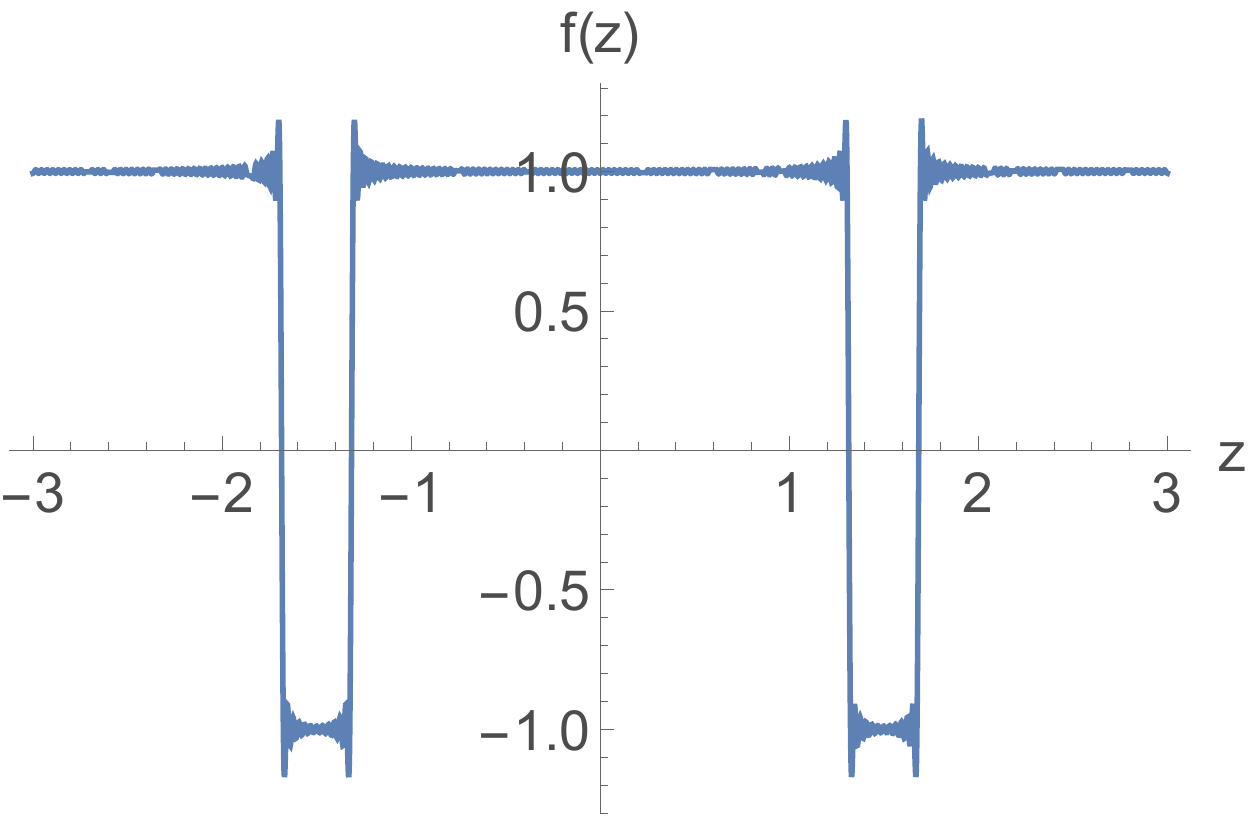}  
    \vspace{-0.4cm}
 
\noindent{{\textit{Figure 2: The Fourier representation of the even periodic mass function we use (100 Fourier terms are used) - in each period it has two domain walls separated by a width $w$.  }}}
\end{center} \vspace{-1cm}

\begin{center}
    \includegraphics[width=9cm]{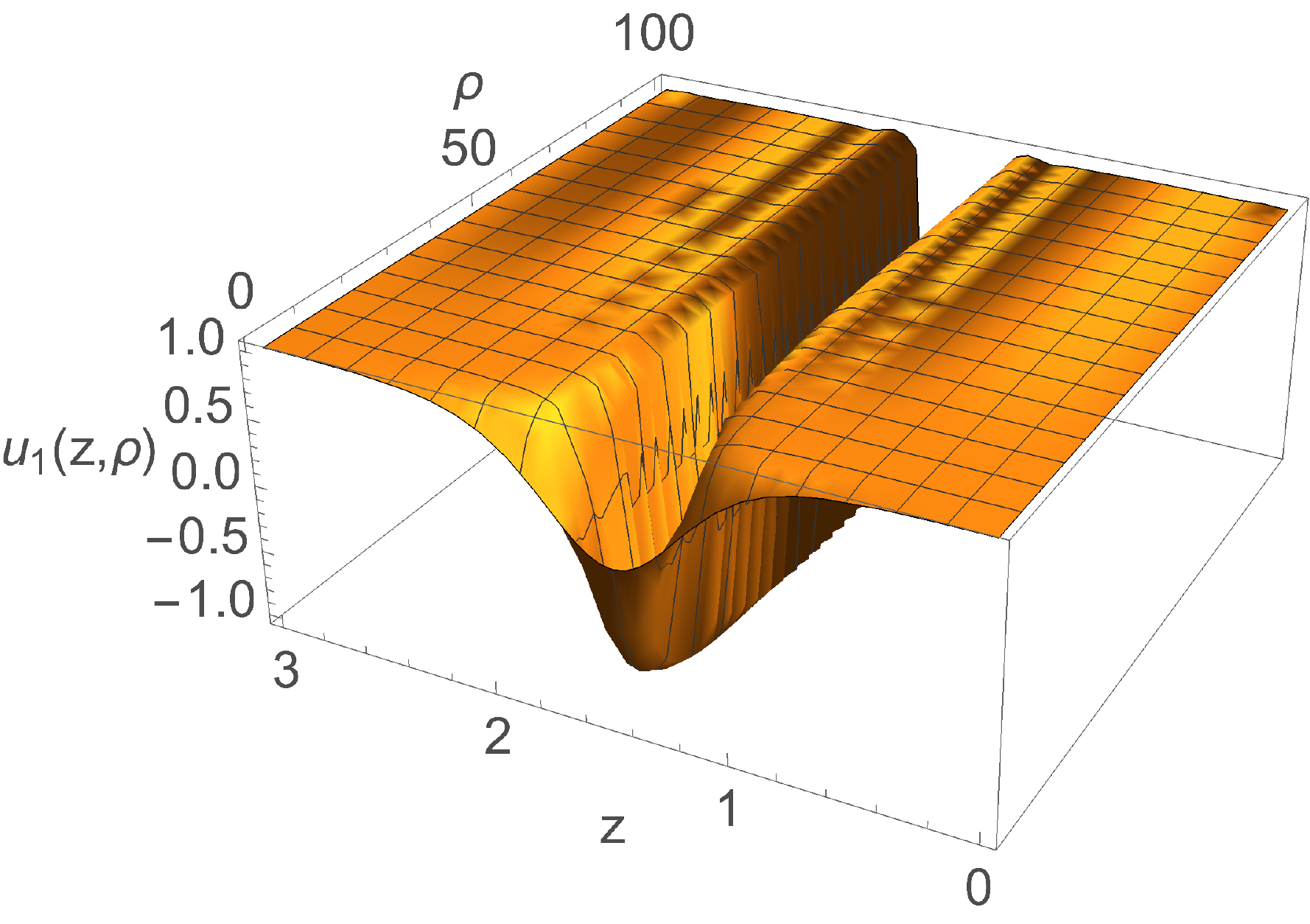}  \vspace{-0.4cm}
    
\noindent{{\textit{Figure 3: the full $\rho-x_3$ dependence of a domain wall pair as represented by (\ref{full})  }}}
\end{center}  \vspace{-0.5cm}

\begin{center}
    \includegraphics[width=8cm]{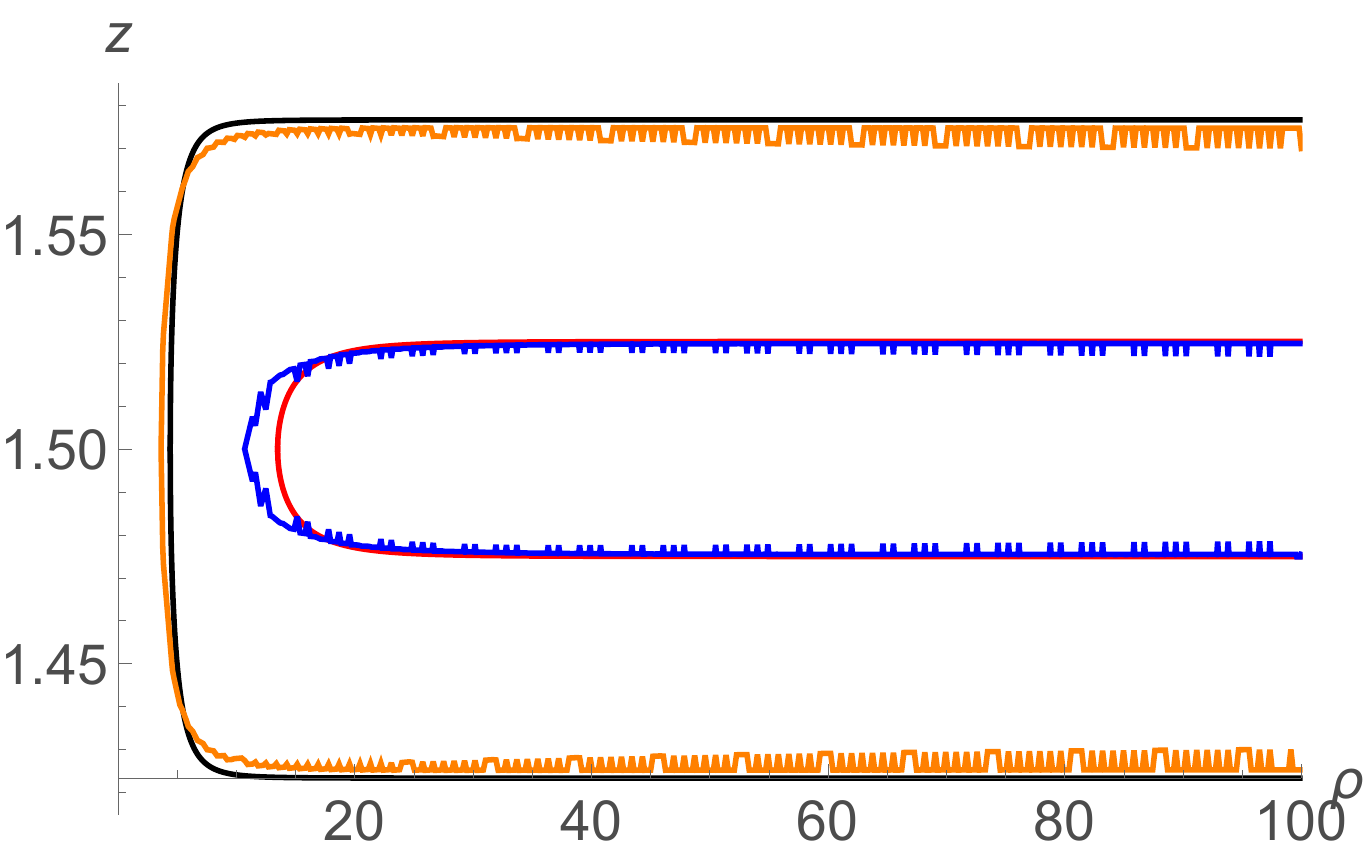}  \vspace{-0.4cm}
    
\noindent{{\textit{Figure 4: The contours in the $\rho-x_3$ plane where $M=0$. The examples given are a numerical solution (Orange), and a second narrower numerical case (Blue), With D5 embeddings from (\ref{d5sol}) of matching width overlaid in (Black) and (Red) respectively. Note the imperfections are due to truncating  the Fourier Series. }}}
\end{center}

\begin{center}
    \includegraphics[width=8.5cm]{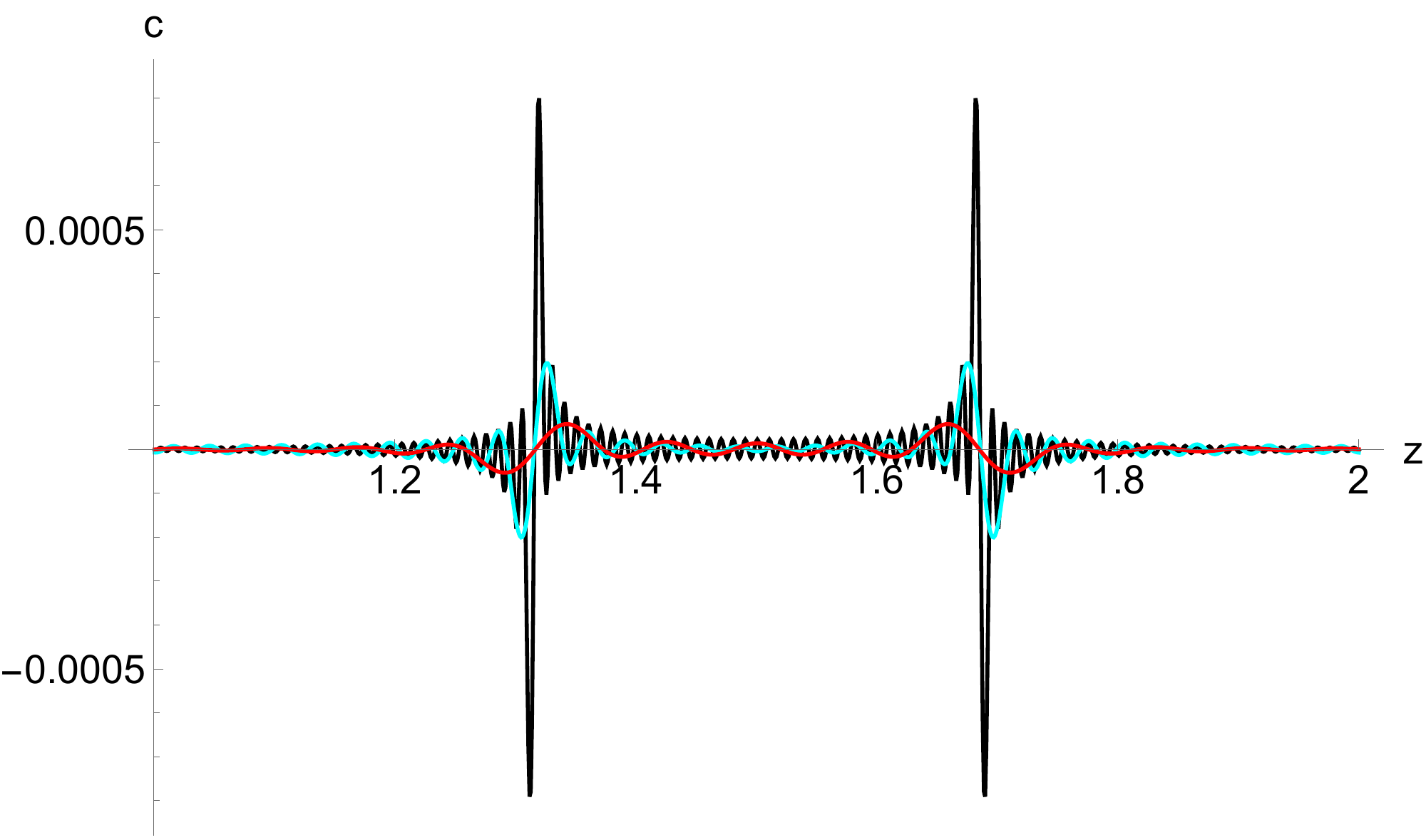}  \vspace{-0.4cm}
    
\noindent{{\textit{Figure 5:  The quark condensate parameter $c$ plotted against $x_3$ across the domain walls for a configuration of width $= 0.37$ and reaching to a depth of $\rho_{min} = 1.27$. First $40$(Red), $100$(Cyan), $300$(Black) Fourier modes included.}}}
\end{center} 

This solution provides the UV boundary data for the holographic field $u_1$. It is now straightforward to plot the 
configuration into the interior of AdS. We simply use our solutions $f_k(\rho)$ as a multiplier on each Fourier mode
\begin{equation} \begin{array}{ccc}
  u_1(\rho,z)& = & \frac{a_0}{2}f_{0}(\rho)+ \sum_{n=1}^{\infty} a_n f_{2 \pi n} (\rho) \cos{\frac{2\pi nz}{3L}} \end{array}. \label{full}
  \end{equation}
  
  We plot an example solution in Figure 3. The high $k$ modes die away as one moves to smaller $\rho$ and the well configuration begins to decay. The key question is where are the contours where $u_1=0$ - this is where the 2+1d fermions will be isolated. We plot this in Figure 4

The two domain walls in the UV are well separated but they join together in the IR. The behaviour of two domain walls joining is very familiar from probe brane embeddings (for example that in section II). As first introduced in the Sakai-Sugimoto  model \cite{Sakai:2004cn}, when branes join in this fashion it indicates condensation of the fermions on the two boundaries. The minimum $\rho$ value of the configuration $\rho_{\rm min}$ is the mass
gap of the theory (formally $\rho_{\rm min}/2 \pi \alpha'$). We will make the same interpretation here. The two initially separated 2+1d 2-spinors have a symmetry breaking interaction together. 
What is not yet clear is whether the symmetry breaking is intrinsic through a mass term or due to spontaneous symmetry breaking.

Of course, strictly the gauge invariant operator that condenses is a path ordered Wilson line stretched between the UV Domain Walls \cite{Aharony:2008an}
\begin{equation} {\cal O} = \bar{q}_1 e^{i \int A_\mu dx^\mu} q_2 \end{equation}
but in the IR at the condensation scale the theory can no longer ``see" the separation (the domain walls  have joined) and the operator will mix freely with the local operator $\bar{q}_1 q_2$. One would expect their vevs to be proportional. We can extract the local 3+1d quark condensate from the sub-leading behaviour of our solution at the 

\begin{center}
    \includegraphics[width=9cm]{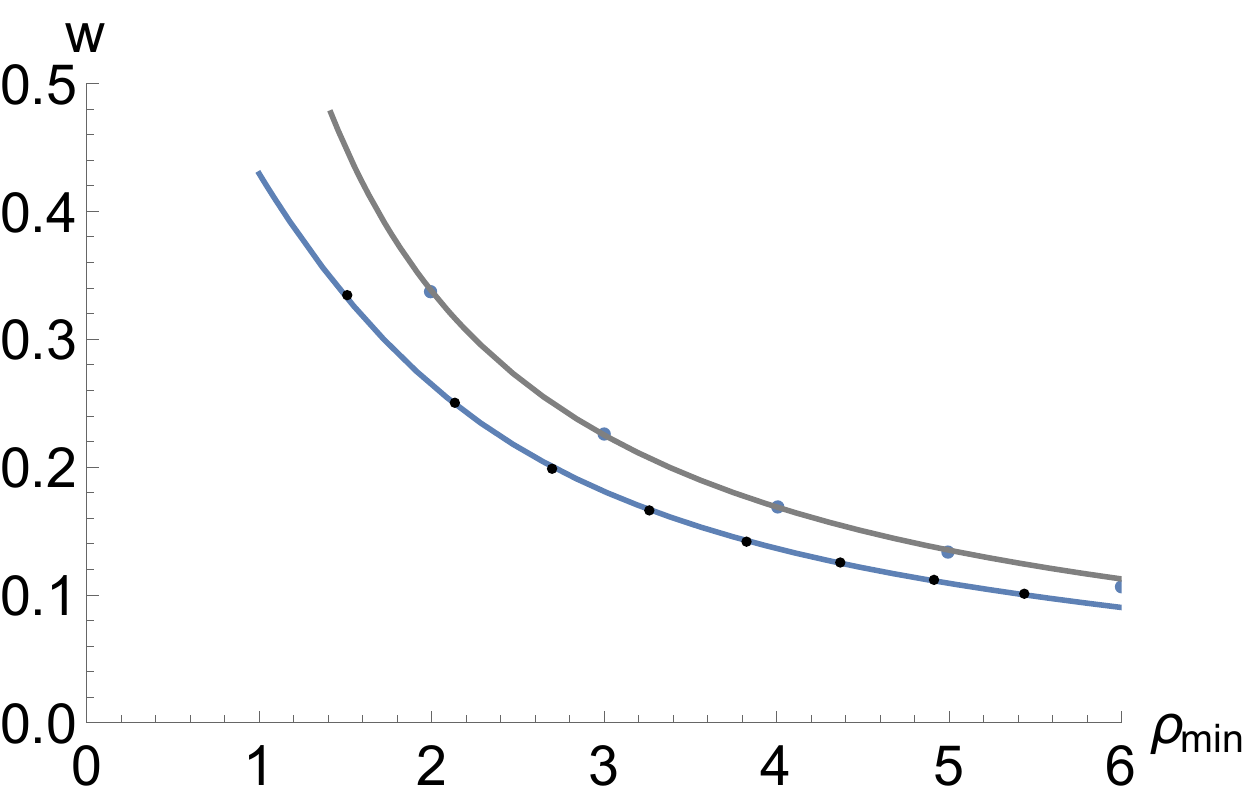}  \vspace{-0.4cm}
    
\noindent{{\textit{Figure 6: The minimum value of $\rho$ an $M=0$ contour reaches to as a function of width between the two domain walls. The D7 Domain wall solution is in blue. The grey is the D3/probe D5 system from Section VI.}}}
\end{center} \vspace{-0.3cm}

boundary (again it falls off as $u_1 \sim m+ c/\rho^2$ with $c$ proportional to the quark condensate). We plot this in Figure 5 where we see the condensate is localised at the defects and becomes more so as one increases the number of Fourier terms. However, note that the solutions have $c=0$ at the domain wall's centre with two peaks, one positive and one negative, to either side that are moving into the domain wall as we increase the number of Fourier modes. Presumably they eventually merge with 
 the true condensate being the sum of the peaks (which could be zero) but this is very hard to compute at the 3+1d level. Below we will restrict our computation to the $M=0$ locus which gives us a better understanding. 

$\rho_{\rm min}$ is the most easily extracted quantity and we can test its dependence on the separation of the domain walls. We 
show this in Figure 6. 
We can fit to the functional form 
\begin{equation} \rho_{\rm min} = \frac{C }{w^p},  \label{guess}  \end{equation}

 For small widths, where the configurations lies well above 
the IR cut off at $\rho=1$, the numerical fit is  $c=0.59 ~p=0.96$. As one expects the relation is governed by dimensional analysis with the separation of the domain walls the only dimensionful parameter in the theory i.e. $p=1$.

At this point it is worth making a harder comparison between these domain wall solutions and the vacuum configuration of the D3/probe D5 $\overline{D5}$ system of Section II. In Figure 4 we have also plotted U-shaped D5 embeddings of the same width as  configurations - they lie very close. In Figure 6 we plot $\rho_{\rm min}$ against the width of the U-shape also. As the domain wall results become more trusted away from the domain wall at $\rho=1$ the two solutions converge. It seems likely from this that the deviations are artefacts of our IR wall. We will prove their equivalence for large quark mass in the next section. In the field theory this equality presumably follows from both systems consisting of massless fermions on the domain walls interacting by the same ${\cal N}=4$ dynamics. The mass gap and self energies of the quarks as a function of energy scale must be the same in each system.

\subsection{The Large Mass Limit}

To move away from the Fourier analysis approximations and numerics we can instead consider two isolated domain walls where the background spatial dependent mass is infinite (or $M \gg 1/w$). That is the mass is strictly zero on the domain wall but infinite elsewhere. In this limit we can derive the contour in the $\rho-z$ plane where the domain wall sits. 

Our solution for $\partial_\rho u_1$ in (\ref{dupe}) will be a delta function on a contour where some $\rho(z)$ vanishes, where $M=0$, multiplied by some very large number, $N$. Keeping just the leading terms in $\partial_\rho u_1$ leaves
\begin{equation} S_{D7} \approx \int d^4x ~d\rho\,\, \rho^3 (\partial_\rho u_i) \sqrt{1+\frac{1}{(\rho^2 + u_i^2)^2}(\partial_z \rho(z))^2} \label{dupe2}
\end{equation}
We must be careful though with the treatment of the metric by the delta function in $(\partial_\rho u_i)$: in particular, a delta function reduces the action to that on a sub-space and so we must correctly adjust the $\sqrt{-g}$ factor to that on the line $\rho(z)$ by including a Jacobian factor. We find it instructive here to consider the problem in a flat 2-plane space where the action would be just
\begin{equation} S \approx \int dz ~d\rho\,\,  (\partial_\rho u_i) \sqrt{1+(\partial_z \rho(z))^2} \end{equation} 
We must set 
\begin{equation} \partial_\rho u_i = {1 \over \partial_z \rho}  \delta(z-z_0)\end{equation}
in order to obtain 
\begin{equation} S \approx \int d\rho\,\,   \sqrt{1+(\partial_\rho z)^2} \end{equation} 
which is the line element on $z(\rho)$. In a curved space this naturally becomes
\begin{equation} \begin{array}{ccl}
\partial_\rho u_i &= & \left. {1 \over \sqrt{g_{\rho \rho } (\partial_z \rho)^2} }   \delta(z-z_0) \right|_{\rm locus} \\ && \\ &=& {\rho \over \partial_z \rho} \delta(z-z_0) \label{delta} \end{array}\end{equation} 
Note both sides of this equation are correctly dimensionless. Equally the pre-factor of the delta function on the right  has dimension of inverse energy so correctly reduces the dimension of the action by one as we move down one in spatial dimension. 

The action (\ref{dupe}) reduces in dimension by one, and writing just the coefficient of the large $N$, gives
\begin{equation} \label{reduced} S = \int d^2x ~d\rho~ \rho^2 \sqrt{1+ \rho^4(\partial_\rho z)^2 } \end{equation} 
The action for $z$ is precisely that of the D3/probe D5 $\overline{D5}$ system (\ref{d5act}) with solution (\ref{d5sol}). Again we see that the dynamics of the systems, mass gaps and so forth are, remarkably, precisely the same. 

\subsection{Fluctuations on the domain wall}

We will now assume that the $M \rightarrow \infty$ limit action (\ref{reduced}) sets the $M=0$ contour to that of (\ref{d5sol}) and any dynamics in the 2+1d theory is a perturbation on this contour ($u \ll M$). 
We can then understand the quark condensate in the system as follows. We start again from the action for $u_1$ (\ref{dupe}) but impose the dynamics we have found by requiring the solution to only lie on the locus in (\ref{d5sol}) by including by hand a delta function of the form in (\ref{delta}). This gives
\begin{equation} {\cal L} \approx {\rho^4 (\partial_\rho z)} \sqrt{1+{\cal A}(\partial_\rho u_i)^2+\frac{(\partial_{x_{2+1}} u_i)^2}{(\rho^2 + u_i^2)^2} }\label{u1locusn}
\end{equation}
with \begin{equation} {\cal A} = 1 +  {1 \over (\partial_\rho z)^2(\rho^2 + u_i^2)^2 }    \ \end{equation}
where from (\ref{fff}) we know
\begin{equation}\partial_\rho z = {\rho_{\rm min}^4 \over \sqrt{\rho^{12} - \rho^8_{\rm min} \rho^4}} \end{equation} 
Note that our theory diverges from the D3/probe D5 $\overline{D5}$ system because the number of scalar fluctuations ($i=1,2$) originates from the D7 probe action. In the field theory this reflects the fact that there is a single 3+1d four component spinor reduced to a single two component spinor on each defect. 

If we consider the vacuum of the theory where there is no $x_{2+1}$ dependence (i.e. $u_1(\rho)$)  we can see by inspection that  (\ref{u1locusn}) is minimized by $\partial_\rho u_1=0$ or $u_1=$ a constant. Equivalently we can see this solution satisfies the equation of motion
\begin{equation} \label{u1no} \begin{array}{l} \partial_\rho \left( {\rho^4 {\cal A} (\partial_\rho z)  \over  \sqrt{1 + {\cal A} (\partial_\rho u_1)^2}} (\partial_\rho u_1) \right) \\ \\
+   {2  \over (\rho^2+u_1^2)^3} {\rho^4 {\cal A}  \over (\partial_\rho z) \sqrt{1 + {\cal A} (\partial_\rho u_1)^2}}  (\partial_\rho u_1)^2 u_1 = 0 \end{array}  \end{equation}

We conclude that for consistency we must fix this constant to be $r_{\rm min}= 0.675/w$, the IR mass gap. That mass is then the same at all RG scales and  there is no condensate in the system. The system  simply describes a massive quark state in a conformal gauge background. 

The $u_2$ field is interesting because it plays the role of the Goldstone boson in systems with chiral symmetry breaking. Here where there is no dynamical chiral symmetry breaking so far,  we don't expect to see Goldstone dynamics. We can write the linearized equation of motion for $u_2(\rho,x)$ on the locus in the background of $u_1$ (thus setting 
$\partial_\rho u_1=0$) 
\begin{equation}   \partial_\rho ( \rho^4 {\cal A} (\partial_\rho z) (\partial_\rho u_2) ) + M^2_{u_2} {\rho^4 (\partial_\rho z)  \over (\rho^2+u_1^2)^2} u_2   = 0  \end{equation}
By rescaling $z, u_2, \rho, M_{u_2}$ we can set $\rho_{\rm min}=1$ in the equation and  therefore for a generic $r_{\rm min}$: 
$M_{u_2} = M_{\rho_{\rm min}=1}/\rho_{\rm min}$. Numerically we find (by shooting from $u_2'(0)=0$ and requiring that $u_2$ vanishes in the UV) that $R^2 M_{\rho_{\rm min}=1}= 7.8$. 

That this 2+1d state is not massless means it is not a Goldstone boson. One has to again conclude, since the theory has a single scale set by the width $w$, that there is a bare quark mass $\rho_{\rm min}$ in the system. Then all bound states naturally have mass proportional to $\rho_{\rm min}$. The joining of the branes is therefore a reflection of the presence of a hard quark mass in this case. 

That the basic domain wall set up has a (non-local) quark mass of $0.675/w$ even in the infinite 3+1d mass, $M$, limit should be compared to the weak coupling domain wall system where the mass is strictly zero in this limit. The extra ingredient is presumably the strong coupling gauge dynamics.

In the next section we will introduce a magnetic field background that is known in some systems to trigger dynamical chiral symmetry breaking and this will lead us to chiral symmetry breaking constructions.

\section{V Dynamical Symmetry Breaking}

The domain wall system we have constructed so far simply describes isolated two component quarks each on a separate 2+1d domain wall. There is a (non-local) mass term linking the quarks of order $1/w$ where $w$ is the separation between the domain walls.  In this section we want to add in dynamics associated with the ${\cal N}=4$ gauge fields that cause chiral symmetry breaking. In fact this system, presumably because the fermions are isolated from each other, are more difficult to condense than those on the usual single probe brane constructions as we will see. 

The expected cause and effect are well known from other systems. If the core of the bulk geometry becomes repulsive to the domain wall (due to a factor growing in the metric as some power of $1/\rho$) then the domain wall will be restricted to lie above some minimum $\rho$ value, $\rho_c$. $\rho_c$ is then interpreted as the chiral symmetry breaking scale and, crucially, as the UV quark mass falls (or equally the separation of the domain wall grows in this case) this scale should remain fixed. We will explore example systems that both realize and fail to realize this phenomena below. 

To begin to explore these issues let's consider the effects of an applied magnetic field which is usually a well controlled source of chiral symmetry breaking.

\subsection{Applied Magnetic Field/Dilaton Profile} 

To include an explicit possible source of dynamical symmetry breaking into the domain wall configuration we will include a magnetic field in the $z$ or $x_3$ direction. Magnetic fields are known to generate chiral symmetry breaking both in field theory \cite{Gusynin:1995nb} and holographic settings \cite{Filev:2007gb}. Our magnetic field  enters as the $1,2$ components of $F_{MN}$ in the DBI action for the probe D7 
\begin{equation} S_{D7} = -T_7\int d^8\xi \sqrt{-det P[G_{MN} + 2 \pi \alpha' F_{MN}]} \end{equation}
which gives  an overall pre-factor on the Lagrangian
\begin{equation} {\cal L}_{D7} \approx  h(r) \rho^3 \sqrt{1+(\partial_\rho u_i)^2+\frac{R^4}{(\rho^2 + u_i^2)^2}(\partial_x u_i)^2} \label{u1Bact}
\end{equation}
with
\begin{equation} h = \sqrt{ 1 + {B^2 R^4 \over (\rho^2 + u_i^2)^2}.   } \label{Bform}\end{equation}
We can effectively set $R=1$ by rescaling $x_{3+1}$ and $B$. 

The magnetic field naturally acts to generate chiral symmetry breaking on the probe D7 brane itself \cite{Filev:2007gb}. This effect destabilizes the linearized discussion in the previous section - the Fourier modes $f(k)$ now satisfy
\begin{equation} \label{u1B}
\partial_\rho\left(h(\rho) \rho^3\partial_\rho u_1\right) - h(\rho) \frac{k^2}{\rho} u_1 + {2 B^2  \over h(\rho) \rho^3}  u_1=0
\end{equation}
The low $k$ modes are unstable and tend to rise to large values on the IR wall. This is not the instability we are hoping to see - we want to watch dynamics in the domain wall theory. To avoid this issue we will therefore move to the large $M$ limit. A very massive quark is insensitive to the IR $B$ field so in the $M \rightarrow \infty$ limit only the domain wall 2+1d locus where $M=0$ will be affected by the magnetic field. 

We therefore start from (\ref{u1Bact})  and take the large $M$ limit with $\partial_z M$ proportional to the delta function in (\ref{delta}). We also assume $u_i=0$, that is that it is much less than $M$. We arrive at the equation for the locus where $M=0$
\begin{equation} \label{reducedB} S = \int d^2x ~d\rho~ h(\rho) \rho^2 \sqrt{1+ \rho^4(\partial_\rho z)^2 } \end{equation} 
There is still a conserved quantity and we obtain
\begin{equation} \label{mip}
\partial_\rho z = {1 \over \sqrt{c^2 h(\rho)^2 \rho^{12} - \rho^4 }}
\end{equation}
where $c$ is the integration constant. The minimum value of $\rho$ a U-shaped configuration reaches is given when the denominator vanishes. If we more generically imagine a function
\begin{equation} h^2 = 1 + {1 \over \rho^{q}} \label{pheno} \end{equation} 
then the vanishing of the denominator in (\ref{mip}) becomes the solution of the polynomial equation
\begin{equation} \label{poly}
    c^2 h^2 \rho^8 - 1=0 \end{equation}
For $q \leq 8$ the polynomial has positive powers of $\rho$ only and vanishes at some $\rho_{\rm min}$ controlled by the constant $c$. By choosing $c$ one can place the zero at any $\rho$. These configurations are U-shaped with the infinite separation case corresponding to $\rho_{\rm min} \rightarrow 0$. Such cases therefore do not display a fixed minimum, $\rho_c$, mass gap as the quark mass falls to zero. They do not describe chiral symmetry breaking. Of course the B field case falls into this category and so does not generate chiral symmetry breaking for the fermions separated on the domain walls.

In contrast to the B-field case, were $q>8$ in (\ref{pheno}) then the polynomial where the denominator of (\ref{mip}) vanishes diverges at both large $\rho$ and as $\rho \rightarrow 0$. Between these limits there is a minimum. For appropriate choices of the constant $c$ the largest $\rho$ root corresponds to the $h=1$ limit. However, as we move in towards smaller $\rho$, eventually, the minimum of the function lifts off from zero and at some fixed $c$ or $\rho_c$ there cease to be further solutions. 
Here we find U-shaped configurations which, as they widen, saturate to falling in no further than $\rho_c$. This is the chiral symmetry breaking effect we were looking for. Clearly we need a rapidly diverging $h$ factor to provide a powerful enough dynamic to trigger chiral symmetry breaking. 

Given that these forms for $h$ (which occurs in the position of the dilaton $e^{-\phi}$ in the action) do trigger chiral symmetry breaking we will briefly study the model with  $q=10$ in (\ref{pheno}). It is not a system we know how to generate in a top-down model but is an interesting toy with 

\begin{center}
    \includegraphics[width=9cm]{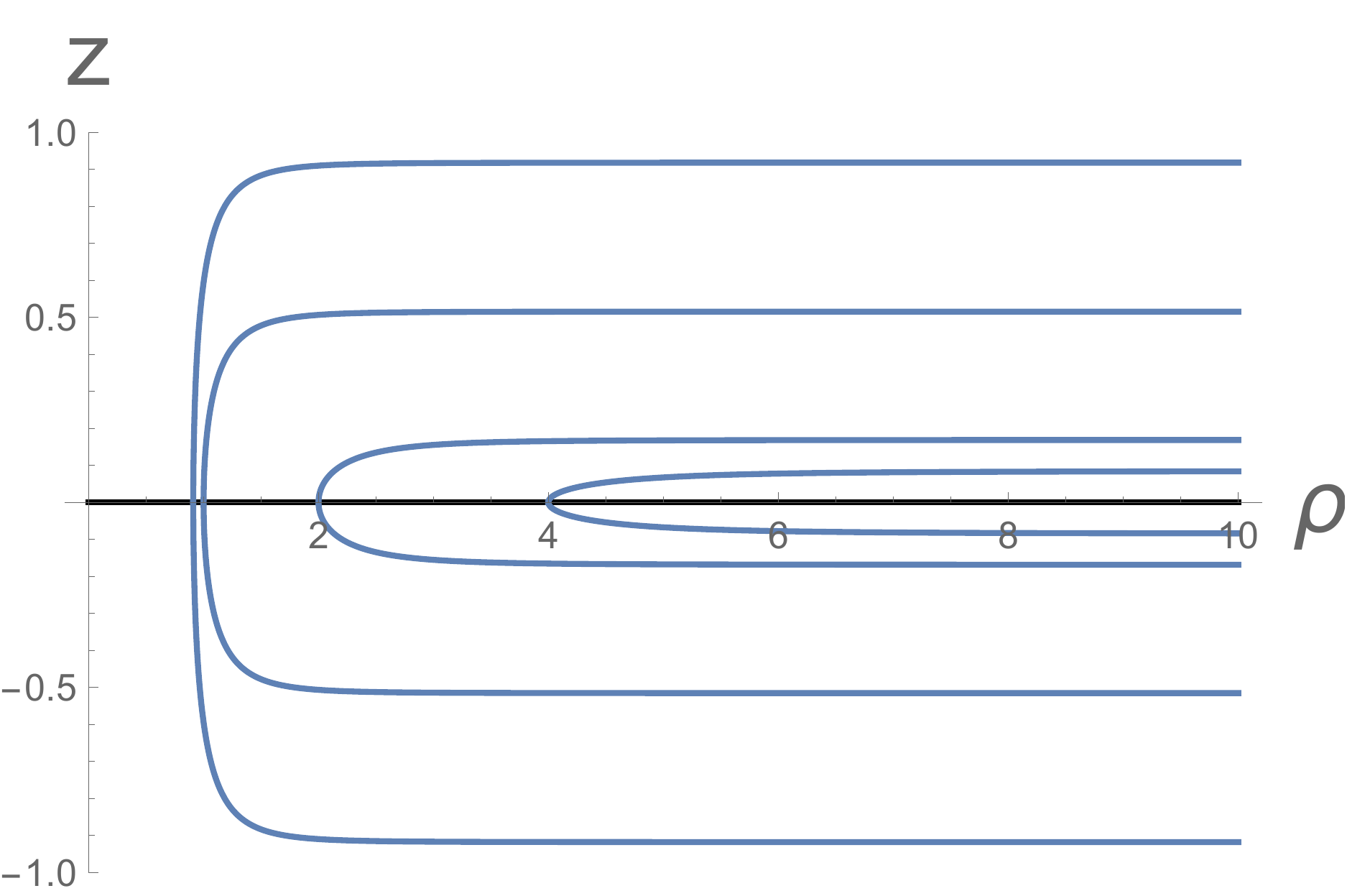}  \vspace{-0.4cm}
    
\noindent{{\textit{Figure 7:  U-shaped loci of the domain wall in the $z-\rho$ plane in the theory with $q=10$ in (\ref{mip}). Note the solutions with large widths pile up at $\rho_c=0.867$ }}}
\end{center} \vspace{-0.3cm}

a phenomenologically imposed (unbackreacted) dilaton profile. In Figure 7 we show the U-shaped loci where $M=0$ for this model displaying the pile up at a fixed $\rho_c$ IR mass scale for widely separated domain walls (with small UV quark mass). 

Again we can determine the behaviour of the sub-leading $u_i$ fields from (\ref{u1Bact}) with (\ref{pheno}) after restricting the dynamics  to the loci in Figure 7 by including by hand a delta  function of the form in (\ref{delta}). This gives
\begin{equation} {\cal L} \approx h(r) {\rho^4 (\partial_\rho z)} \sqrt{1+{\cal F}(\partial_\rho u_i)^2+\frac{(\partial_{x_{2+1}} u_i)^2}{(\rho^2 + u_i^2)^2} }\label{u1locus}
\end{equation}
with \begin{equation} {\cal F} = 1 +  {1 \over (\partial_\rho z)^2(\rho^2 + u_i^2)^2 }    \ \end{equation}
where $\partial_\rho z$ is given in (\ref{mip}).

The $u_1$ vacuum equation is
\begin{equation}  \label{u1eom} \begin{array}{l} \partial_\rho \left( {h ~\rho^4 {\cal F} (\partial_\rho z) \over  \sqrt{1 + {\cal F} (\partial_\rho u_1)^2}} (\partial_\rho u_1) \right) \\ \\
+   {2  \over (\rho^2+u_1^2)^3} {h~ \rho^4  \over  (\partial_\rho z) \sqrt{1 + {\cal F} (\partial_\rho u_1)^2}}  (\partial_\rho u_1)^2 u_1 \\ \\
- 2{\partial h \over \partial r^2} {\rho^4 (\partial_\rho z)} \sqrt{1+{\cal F}(\partial_\rho u_1)^2 }~u_1  = 0 \end{array}  \end{equation}
The extra term relative to (\ref{u1no}), due to $h$, if  sufficiently large, can cause condensation. Note the mechanism here is the same as discussed for D7 probe examples in \cite{Alvares:2012kr} - the final term can be considered a running mass for $u_1$ and if it violates the Breitenlohner Freedman (BF) bound  \cite{Breitenlohner:1982jf} at some $\rho$ then the $u_1=0$ solution becomes unstable. 

\begin{center}
    \includegraphics[width=9cm]{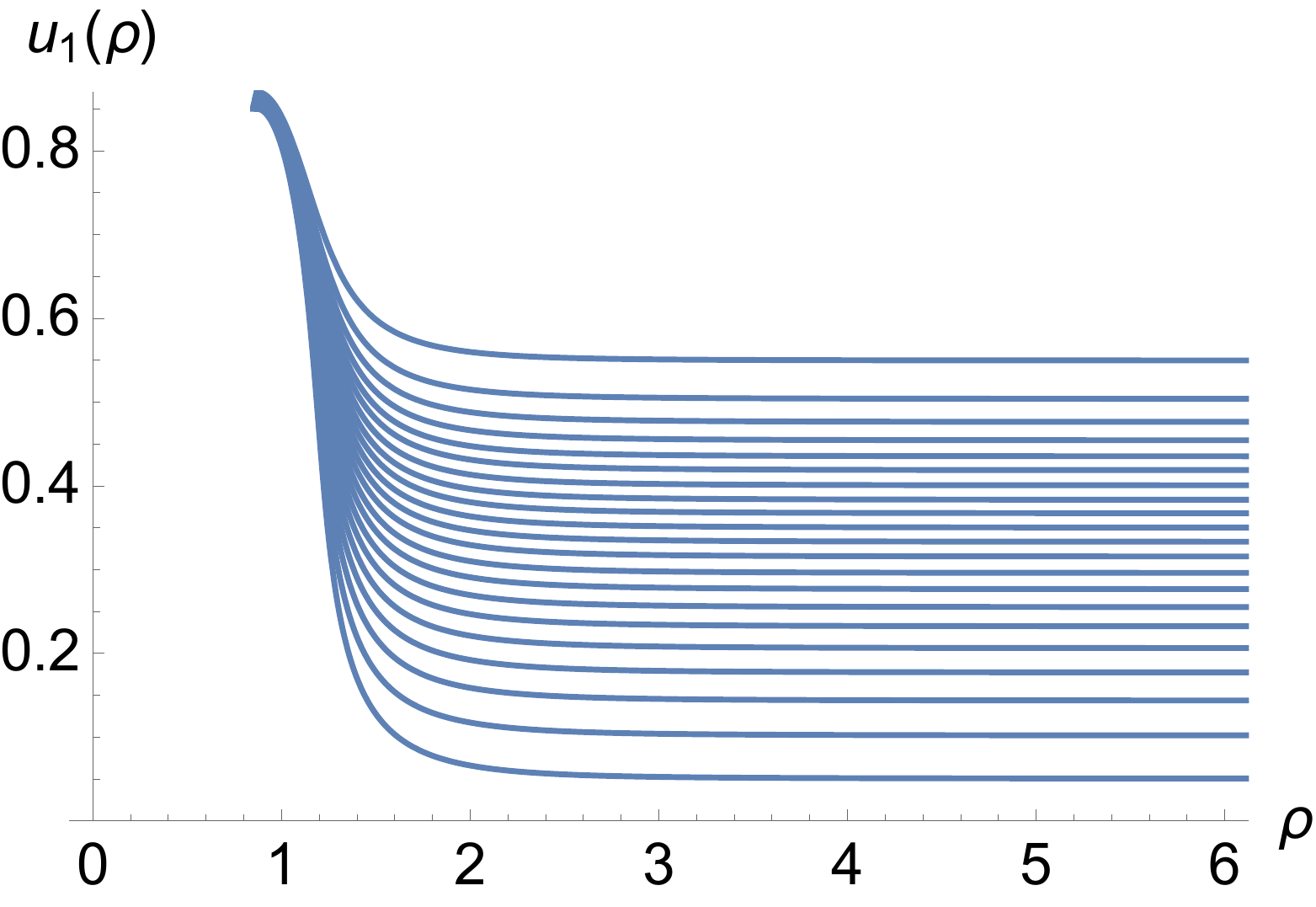}  \vspace{-0.4cm}
    
\noindent{{\textit{Figure 8: the vacuum functions $u_1(\rho)$ for the theory with $q=10$ in (\ref{mip}) showing chiral symmetry breaking behaviour. Note the solutions begin at $\rho_{\rm min}$ in the IR.}}}
\end{center} \vspace{0.2cm}

We solve (\ref{poly}) to set $c$ for a given $\rho_{\rm min}$ (this involves more fine tuning the closer the U-shape approaches $\rho_{\rm min}$ and the separation of the domain walls goes to zero). We then solve (\ref{u1eom}) subject to $u_1(\rho_{\rm min})=\rho_{\rm min}$ and $u_1'(\rho_{\rm min})=0$ for different $\rho_{\rm min}$. The results are shown in Figure 8. They show clear chiral symmetry breaking behaviour with the IR mass becoming independent of the UV mass at small UV mass. 

Note we have also checked examples where $q<8$ and there the extra term in the equation of motion for $u_1$ 
does not violate the BF bound and the IR mass approaches  zero with the UV mass.  This is self consistent with the 
loci shape in these theories which do not show chiral symmetry breaking.

Finally we can write the linearized equation of motion for $x$-dependent $u_2$ fluctuations in the $u_1$ background
\begin{equation}  \label{u2eom} \begin{array}{r} \partial_\rho \left( {h ~\rho^4 {\cal F} (\partial_\rho z) \over  \sqrt{1 + {\cal F} (\partial_\rho u_1)^2}} (\partial_\rho u_2) \right) + M_{u_2}^2  {h ~\rho^4  (\partial_\rho z) \over  \sqrt{1 + {\cal F} (\partial_\rho u_1)^2} (\rho^2+u_1^2)^2} u_2\\ \\
+   {2  \over (\rho^2+u_1^2)^3} {h~ \rho^4   \over (\partial_\rho z) \sqrt{1 + {\cal F} (\partial_\rho u_1)^2}}  (\partial_\rho u_1)^2 u_2 \\ \\
- 2{\partial h \over \partial r^2} {\rho^4 (\partial_\rho z)} \sqrt{1+{\cal F}(\partial_\rho u_1)^2 }~u_2 = 0 \end{array}  \end{equation}
which can be solved subject to boundary conditions $u_2'(\rho_{\rm min})=0$ and in the UV 
$u_2= 0$ (so the fluctuation is only of the operator and not the source). In the case where the UV solution for $u_1$ asymptotes to zero we can immediately see that this ``pion'' is massless - if we set $M^2=0$ in (\ref{u2eom}) then there is the solution $u_2 \propto u_1$ since then (\ref{u2eom}) becomes precisely (\ref{u1eom}). Since this solution falls to zero in the UV it is appropriate for the massless pion state. At other values of UV quark mass we must 

\begin{center}
    \includegraphics[width=9cm]{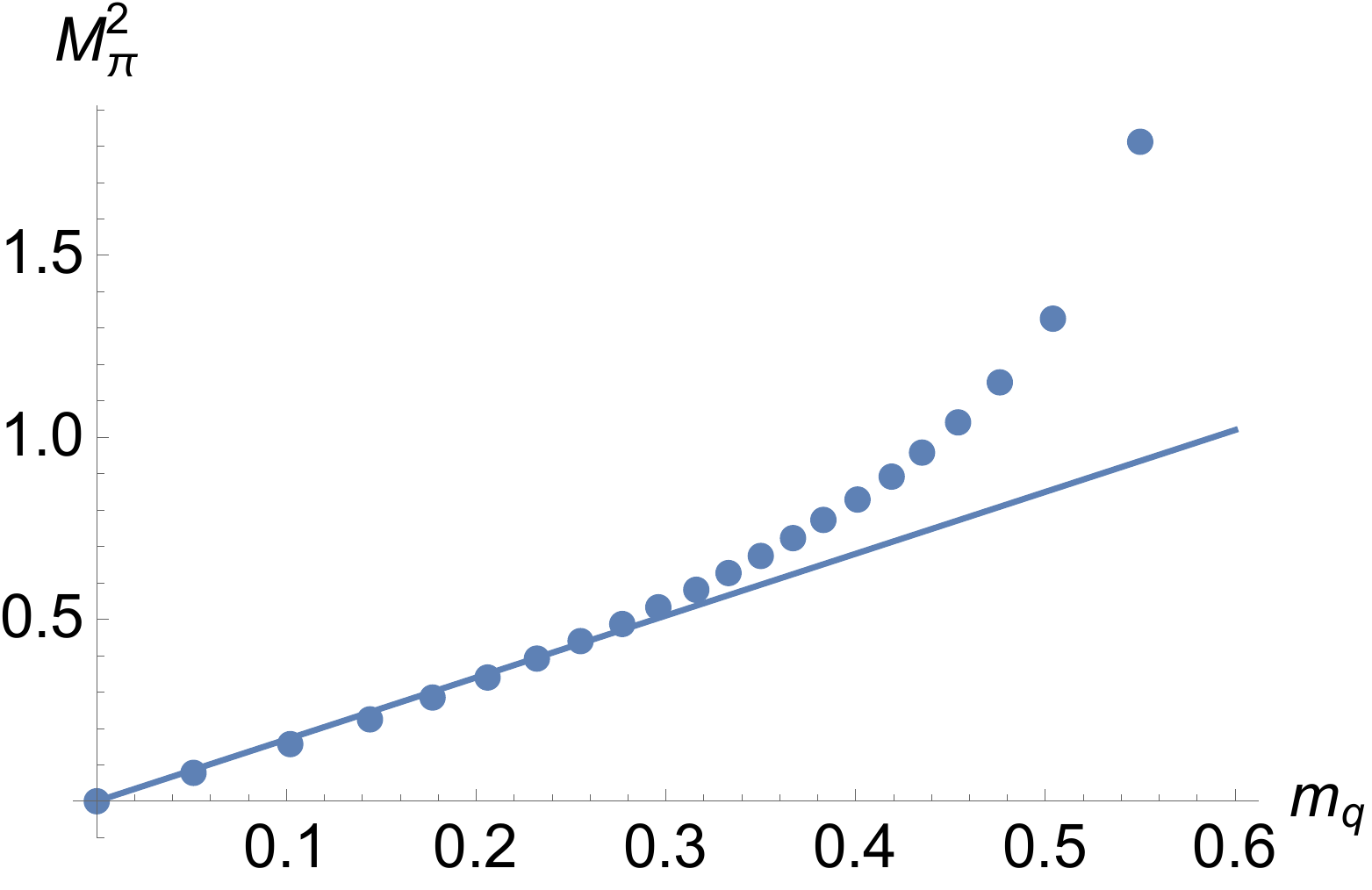}  
    
\noindent{{\textit{Figure 9: For the vacuum solutions in Figure 8: the pion ($u_2$) mass squared against the quark mass (extracted from the UV of Figure 8). Computed data points are shown as well as a guiding linear function. A Gell-Mann Oakes Renner relation is reproduced at small $m_q$ but the system returns to $M_\pi^2 \sim m_q^2$ at larger $m_q$.}}}
\end{center} \vspace{0.1cm}

solve numerically and we plot, as the points in Figure 9, this field's mass squared against the UV quark mass extracted from the solutions in Figure 8. We also provide a linear line to guide the eye. At small $m_q$ the data reasonably suggest a linear Gell-Mann-Oakes-Renner relation - the state is the Goldstone boson (pion) of the symmetry breaking. At larger $m_q$ the relation returns to the expected $M_\pi^2 \propto m_q^2$.

\subsection{Dilaton Flow Geometries}

As another example of a chiral symmetry breaking mechanism we will turn to a backreacted hard wall model. The simplest example is a case of a dilaton flow deformation of AdS. First let's consider the metric from  \cite{Constable:1999ch} (it generates chiral symmetry breaking in the massless D3/probe D7 system as described in \cite{Babington:2003vm}). In {\it Einstein frame} the metric can be written as \cite{Babington:2003vm}
\beq ds^2 = G_{x}\,dx_4 ^2 + G_r ( d\rho^2 + \rho^2 \Omega_3^2 + du_1^2 + du_2^2), \eeq where
\beq G_x = H^{-1/2} \left( { r^4 + b^4 \over r^4-b^4}
\right)^{\delta/4} \eeq and 
\beq G_r = H^{1/2} \left( {r^4 + b^4 \over r^4-
b^4}\right)^{(2-\delta)/4} {r^4 - b^4 \over r^4 } \eeq
with
\beq
H =  \left(  { r^4
+ b^4 \over r^4 - b^4}\right)^{\delta} - 1.
\eeq
Here $\Delta^2 + \delta^2 = 10$ and $\delta = L^2/2$. The dilaton is given by
\beq e^{2 \phi} = e^{2 \phi_0} \left( { r^4 + b^4 \over
r^4 - b^4} \right)^{\Delta}\eeq
Note here again the radial directions are $r^2=\rho^2+u_1^2 + u_2^2$. 
 The geometry has a running coupling growing into the IR but also a singularity at $b$ which it is not clear how to resolve in the full string theory. Nevertheless the singularity is repulsive to probe branes and triggers chiral symmetry breaking in the D3/probe D7 system \cite{Babington:2003vm}. We will use this geometry to trigger chiral symmetry breaking on the domain walls assuming it captures some aspects of a more complete system.  Note that for numerical work one can rescale $\rho,u_i$ to set $b=1$ - it sets the energy scale of the geometry/dual. 

The probe D7 Lagrangian in this geometry is given by 
\begin{equation}  {\cal L}_{D7} =  e^\phi G_x^2 G_r^2 \rho^3\sqrt{1+(\partial_\rho u_i)^2+\frac{G_r}{G_x}(\partial_x u_i)^2}
    \end{equation} 
If the 3+1d theory's quark mass is set to be less than or of order the scale $b$ then the background D7 probe bends off axis and breaks chiral symmetry in the 3+1d theory \cite{Babington:2003vm}. We will therefore again take the $M\rightarrow \infty$ limit so that the 3+1d theory does not have spontaneous breaking, but allow domain walls where $M=0$. Thus we impose that the mass vanishes on a contour $z(\rho)$ by setting
\begin{equation} \label{dpfactor2}
      \partial_\rho u_1 = N \left.\frac{G_r^{-1/2}}{   {\partial_z\rho}}\right|_{z=z_0}\delta(z-z_0)
\end{equation}
and keeping the terms leading in $N$. We obtain the Lagrangian for the locus $z(\rho)$
\begin{equation}  {\cal L}_{D7} =  e^\phi G_x^{3/2} G_r^{2} \rho^3\sqrt{1+ \frac{G_x}{G_r} (\partial_\rho z)^2}
    \end{equation} 
Note a good cross-check on this result is that it matches the embedding action for a 6-brane placed in the 0-2,$\rho$, and $\Omega_3$ directions with some profile $z(\rho)$. 

There remains a conserved quantity so we find
\begin{equation} 
    \partial_\rho z =\frac{G_r^{1/2}}{G_x^{1/2}\sqrt{c^2 e^{2 \phi} G_x^4 G_r^3 \rho^6-1}}
\end{equation}
The denominator factor in the square root blows up as $\rho \rightarrow \infty, $ and thus if $c$ is too large there are no roots. This means the U-shaped embeddings end at a fixed $c$ or $\rho_{min}$ which is consistent with chiral symmetry breaking as in our previous example shown in Figure 7.

The fluctuations on the brane are described, after the imposition of a delta function to restrict their behaviour to the locus $z(\rho)$ by 
\begin{equation}  \begin{array}{ccl} {\cal L}_{D7} &=&  e^\phi G_x^2 G_r^{2}(r) G_r(\rho)^{-1/2}{\rho^3 \over \partial_z \rho}\\ &&\\ &&\sqrt{1+{\cal F}(\partial_\rho u_i)^2+\frac{G_r}{G_x}(\partial_x u_i)^2}
  \end{array}  \end{equation} 
\begin{equation} {\cal F} = \bigg(1+\frac{G_r}{G_x (\partial_\rho z)^2 }\bigg)\end{equation}
It's worth stressing here that the factors of $G_u(\rho)$ come from (\ref{dpfactor2}) - in (\ref{dpfactor2}) and the analysis below we fixed the loci of the $M=0$ domain wall assuming $u_i=0$. Here we enforce that contour before solving for $u_i$.

The equations of motion for the vacuum for $u_1$ and fluctuations about that vacuum for $u_2$ again follow straightforwardly. 
For the vacuum
\begin{equation}  \label{CM_u1eom} \begin{array}{l}
    \partial_\rho\bigg(e^{\phi}{G_x^2G_r^2\over \sqrt{G_r(\rho)}}\rho^3\partial_\rho z {{\cal F}\;\partial_\rho u_1 \over \sqrt{1+{\cal F}(\partial_\rho u_1)^2}}\bigg)\\ \\ - \Bigg({2\rho^3 \,\partial_\rho z \over \sqrt{G_r(\rho)}} \sqrt{1+{\cal F}(\partial_\rho u_1)^2} \cdot{\partial (e^\phi G_x^2 G_r^2)\over \partial (r^2)} \cdot u_1\Bigg)\\ \\
    -\Bigg(\rho^3 \;\partial_\rho z\; {e^\phi\, G_x^2 \,G_r^2 \;(\partial_\rho u_1)^2 \over \sqrt{G_r(\rho)}\sqrt{1+{\cal F}(\partial_\rho u_1)^2}} \cdot{\partial{\cal F}\over \partial r^2} \cdot u_1 \Bigg)=0
\end{array}\end{equation}
and for the fluctuation,
\begin{equation}  \label{CM_u2eom} \begin{array}{l}
    \partial_\rho\bigg(e^{\phi}{G_x^2 G_r^2\over \sqrt{G_r(\rho)}}\rho^3\partial_\rho z {{\cal F}\;\partial_\rho u_2 \over \sqrt{1+{\cal F}(\partial_\rho u_1)^2}}\bigg)\\ \\ - \Bigg({2\rho^3 \,\partial_\rho z \over \sqrt{G_r(\rho)}} \sqrt{1+{\cal F}(\partial_\rho u_1)^2} \cdot{\partial (e^\phi G_x^2 G_r^2)\over \partial (r^2)} \cdot u_2\Bigg)\\ \\
    -\Bigg(\rho^3 \;\partial_\rho z\;{ e^\phi\, G_x^2 \,G_r^2 \;(\partial_\rho u_1)^2 \over \sqrt{G_r(\rho)}\sqrt{1+{\cal F}(\partial_\rho u_1)^2}} \cdot{\partial{\cal F}\over \partial r^2} \cdot u_2 \Bigg)\\ \\
    +e^\phi {G_x G_r^3\over\sqrt{G_r(\rho)}}\rho^3 \partial_\rho z {M_{u_2}^2 \, u_2 \over \sqrt{1+{\cal F}(\partial_\rho u_1)^2}}=0.
\end{array}\end{equation}

In Figure 10 we plot the vacuum solutions which again show chiral symmetry breaking as we saw in the case above. In Figure 11 we plot the pion mass squared against the quark mass extracted from the solutions in Figure 10 to again show Gell-Mann-Oakes-Renner behaviour. The backreacted dilaton flow therefore shares the behaviour of the previously considered unbackreacted example.

\begin{center}
    \includegraphics[width=9cm]{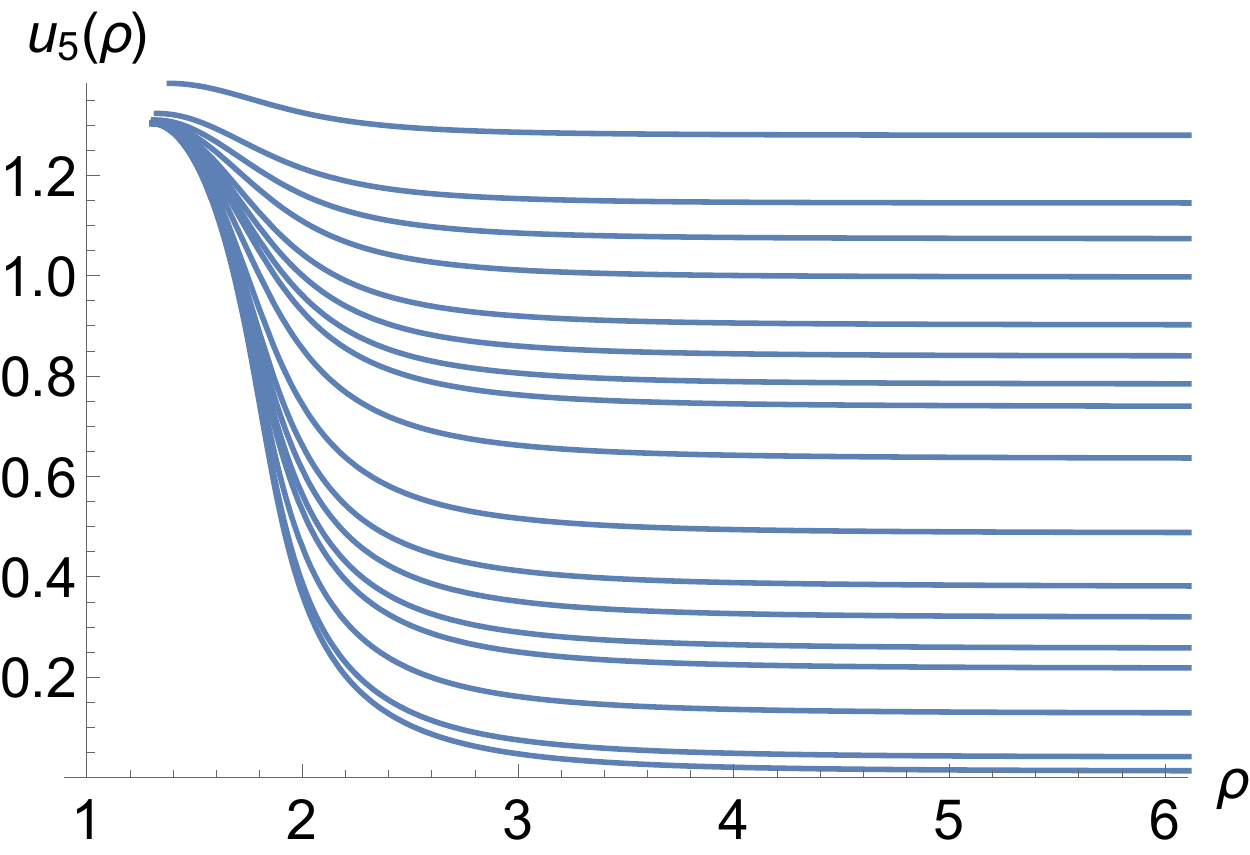}  \vspace{-0.4cm}
    
\noindent{{\textit{Figure 10: the vacuum functions $u_1(\rho)$ for the theory with a dilaton flow in (54)-(58) showing chiral symmetry breaking behaviour. Note the solutions begin at $\rho_{\rm min}$ in the IR.}}}
\end{center}

\begin{center}
    \includegraphics[width=9cm]{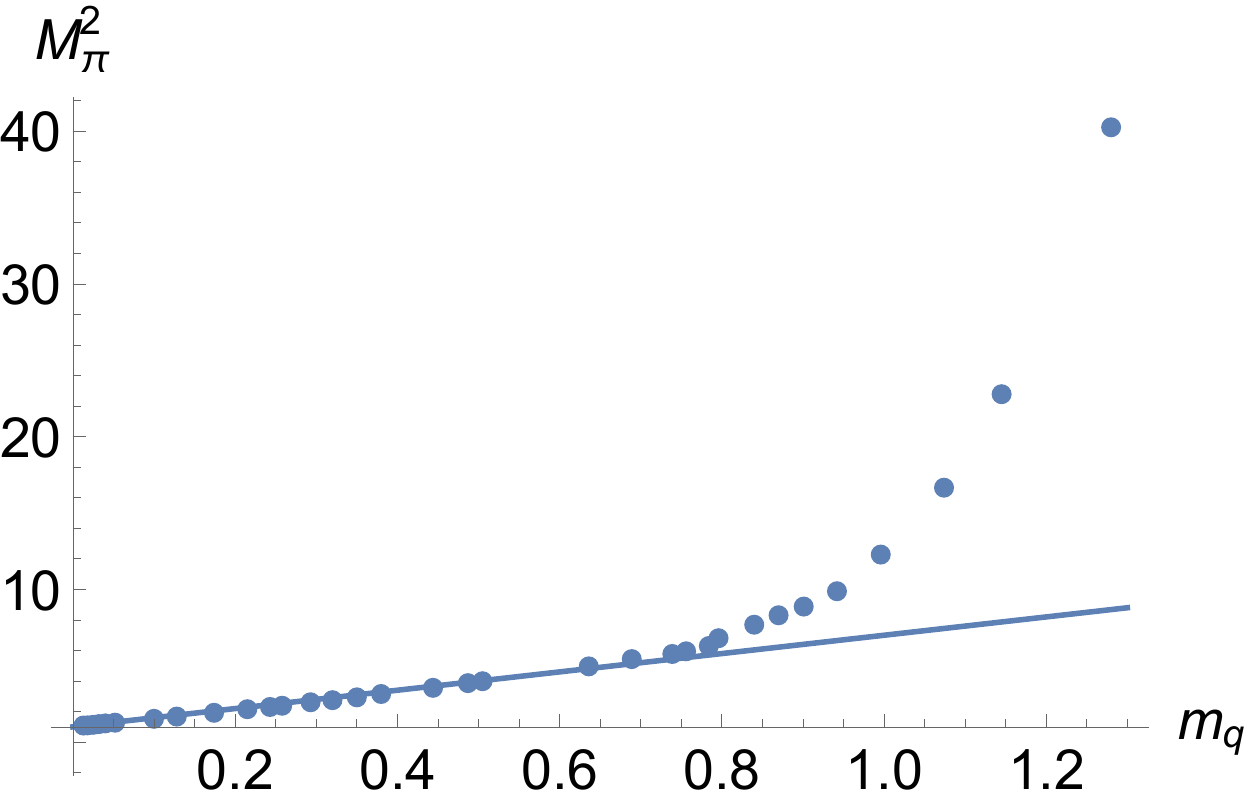}  
    
\noindent{{\textit{Figure 11: For the vacuum solutions in Figure 10, the pion ($u_2$) mass squared against the quark mass (extracted from the UV of Figure 10). Computed data points are shown as well as a guiding linear function. A Gell-Mann Oakes Renner relation is reproduced at small $m_q$ but the system returns to $M_\pi^2 \sim m_q^2$ at larger $m_q$.}}}
\end{center}

We note that there is an alternative dilaton flow geometry presented in  \cite{Kehagias:1999iy} which is (in string frame) 
\begin{equation} \begin{array}{cl}ds_{10}^2 = & e^{\phi/2}\left( \frac{r^2}{R^2}A^2(r)\eta_{\mu\nu}dx^\mu dx^\nu \right.\\& \\ &\left. + \frac{R^2}{r^2}\Big( d\rho^2 + \rho^2d\Omega_{3}^2+ dX_8^2 + dX_9^2\Big)\right)  \end{array} \label{dilflow} \end{equation}
with
\begin{equation} A(r)=\left(1-\left(\frac{r_0}{r}\right)^8\right)^{1/4}  \end{equation}
\begin{equation} e^\phi = \left(  \frac{(r/r_0)^4+1}{(r/r_0)^4-1} \right)^{\sqrt{3/2}}  \end{equation}
Again for a massless D7 the 3+1d theory displays chiral symmetry breaking  \cite{Ghoroku:2004sp}. 
The probe D7 Lagrangian in this geometry is given by 
\begin{equation}  {\cal L}_{D7} =  e^\phi A^4\rho^3\sqrt{1+(\partial_\rho u_i)^2+\frac{1}{A^2 r^4}(\partial_x u_i)^2}
    \end{equation} 
We take the $M\rightarrow \infty$ limit so that the 3+1d theory does not have spontaneous breaking, but allow domain walls where $M=0$. Thus we impose that the mass vanishes on a contour $z(\rho)$ by setting
\begin{equation}
      \partial_\rho u_1 = N \frac{ e^{-\phi/4}\rho}{  {\partial_z\rho}|_{z=z_0}}\delta(z-z_0)
\end{equation}
and keeping the terms leading in $N$. We obtain the Lagrangian for the locus $z(\rho)$
\begin{equation} \label{dpfactor}
{\cal L} =  e^{3\phi/4}A^3 \rho^2 \sqrt{1+A^2\rho^4 (\partial_\rho z)^2}
\end{equation}
The conserved quantity leads to
\begin{equation}
    \partial_\rho z =\frac{1}{A\rho^2\sqrt{c^2e^{3\phi/2}A^{8}\rho^{8}-1}}
\end{equation}
Here though the factor in the denominator does not lead to pile up behaviour and we do not find chiral symmetry breaking. The reason for this distinction is unclear and we do not have a good field theoretic understanding, although presumably this singularity is simply not strong enough to overcome the separation of the chiral fermions to cause condensation.

\section{VI Nambu-Jona-Lasinio Interactions}

It is worth visiting Witten's multi-trace prescription \cite{Witten:2001ua}  in this context. The prescription is that any vacuum configuration with a source present has a second interpretation. The source is viewed as not being intrinsic 
(or a bare Lagrangian term) but resulting from the condensation of the associated operator and the presence of, say, a double trace operator. This mechanism has been explored for Nambu-Jona-Lasinio  operators \cite{Nambu:1961tp} in the 
D3/probe D7 system in \cite{Evans:2016yas}.  

In these domain wall set ups we have a UV mass when the domain  walls are only separated by a finite distance. Again we assume any non-local operators mix wit the local operators described on the domain wall. 
Here we can also consider the UV mass to be due to the condensation of ${\cal O} = \langle \bar{q}_1 q_2 \rangle $ and the presence of the NJL operator
\begin{equation} \Delta {\cal L} = {g^2 \over \Lambda_{UV}^2}   \bar{{\cal O}} {\cal O}~~~~ \rightarrow ~~~~{g^2 \over \Lambda_{UV}^2}   \langle \bar{{\cal O}} \rangle {\cal O} = m {\cal O}
\end{equation}
In this case the boundary conditions for fluctuations such as the $u_2$ pion should be changed to allow fluctuations in the effective mass term (since it is a reflection of the operator, which can fluctuate). In particular since the mass generation is entirely dynamical one expects a Goldstone mode with $M^2=0$ -  the U(1)$\times$U(1) symmetry on the two 2+1d quarks is broken to the vector U(1). 

For example, in the case of the unbackreacted dilaton profile section above, one interprets the UV mass in the $u_1$ vacuum solutions of (\ref{u1eom}) in Figure 8 to be due to the NJL interaction - reading off the constants $m,c$ in the UV allows one to compute the NJL coupling ($g^2 = \Lambda_{UV}^2 m / c$). Further we can see that if we set $M^2=0$ in the pion's equation (\ref{u2eom}) then there is a solution where $u_2$ is proportional to the vacuum  $u_1$ solution - this follows because with these substitutions (\ref{u2eom}) is identical to (\ref{u1eom}). That solution for $u_2$, by default, asymptotes to the same UV boundary $m,c$ as the $u_1$ background and hence is consistent with the background's NJL interaction $g^2$.  The same logic follows for the back-reacted dilaton profile where in the $M^2=0$ and $u_2 \propto u_1$ limit, (\ref{CM_u2eom}) becomes degenerate with (\ref{CM_u1eom}). 

This logic can also be followed in the basic domain wall example in pure AdS without chiral symmetry breaking but since those solutions have just $u_1=\rho_{\rm min}$, a constant, $c=0$ and $g^2 \rightarrow \infty$. This is similar to the 3+1d supersymmetric D7 probe configuration with an NJL term as described in \cite{Evans:2016yas}.

As an aside here we note that 10 years or so ago there was some discussion in the literature \cite{Antonyan:2006vw,Evans:2007jr} as to whether the Sakai Sugimoto construction, with non-anti-podal D8 branes on the compact direction of the geometry there, described a massive or a NJL induced mass configuration. In the light of the discussion here we now see that this dispute was simply semantics - a vacuum configuration describes both possibilities with the difference simply due to which boundary conditions are imposed on the possible Goldstone mode. 

\section{VII Domain Walls on D5 Probes}

Here we briefly note that we could have presented an equivalent discussion to that above but starting with a single D5 brane probe in the supersymmetric configuration of section II. Here we do not mean the U-shaped configurations of (\ref{d5sol}) but instead just the flat embedding $u_{i}=$ constant that are solutions following from the action
\begin{equation} S_{D5} \approx \int d^4x ~d\rho\,\, \rho^2 \sqrt{1+(\partial_\rho u_i)^2+\frac{R^4}{(\rho^2 + u_i^2)^2}(\partial_x u_i)^2} 
\end{equation}
here $i=1..3$ associated with the $x_7,x_8,x_9$ directions.
These solutions (which asymptote to $m + c/\rho$) describe a single 2+1d defect with ${\cal N}=2$ quark multiplets.  

One can now introduce domain walls that have $M=0$ on a 1+1d slice. Here the expectation is that 1+1d truly chiral fermions are localized at the domain wall boundaries. In the large mass limit we set again
\begin{equation} \partial_\rho u_1 =  N {\rho \over (\partial_z \rho|_{z_0}) } \delta (z-z_0) \end{equation}
so the action reduces in dimension by one, and writing just the coefficient of the large $N$, gives
\begin{equation}  S = \int d^2x ~d\rho~ \rho \sqrt{1+ \rho^4(\partial_\rho z)^2 } \end{equation} 
This is precisely the action of a  D3 probe in the configuration
\begin{center}
 \begin{tabular}{c|c c c c c c c c c c}
      &0&1&2&3&4&5&6&7&8&9   \\   \hline
      D3&-&-&-&-& $\bullet$ & $\bullet$ &$\bullet$ &$\bullet$ &$\bullet$ &$\bullet$ \\
      D3&-&-&$\bullet$&$\bullet$&-&-&$\bullet$&$\bullet$&$\bullet$ &$\bullet$  
 \end{tabular}
\end{center}
\vspace{-1.5cm}

\begin{equation} \end{equation} 

There are again U-shaped domain wall configurations consistent with a UV mass that scales as $1/w$ where $w$ is the separation of the domain walls. One can again introduce a dilaton factor 
\begin{equation} h^2 = 1 + {1 \over \rho^q} \end{equation}
to induce chiral symmetry breaking and the equivalent of (\ref{poly}) is 
\begin{equation}  
    c^2 h^2 r^6 - 1=0 \end{equation}
here one needs $q>6$ to see chiral symmetry breaking so again an applied baryon number  magnetic field does not cause chiral symmetry breaking. The chiral symmetry breaking behaviours for the cases $q>6$ or for the dilaton flow geometry are simple to compute and follow our analysis above. There is no new behaviour so we don't reproduce this case in detail.

\section{VIII Conclusions}

We have explored the localization of fermions on lower dimension domain wall defects in a holographic framework. In the D3/D7 system a step function in the quark mass localizes 2+1 dimensional spinors on the domain wall. We have explored the condensation between these fermions as a result of bulk dynamics. The linearized approximation is useful to understand that the $M=0$ locus forms a U-shaped system as the two domain walls merge in the interior of AdS. In the large mass limit we have solved for this locus and shown it is the same that D5 probes in AdS form. By restricting the D7 fields to the large mass locus we have realized an explicit description of the mass/chiral condensate operator/source pair. When the domain walls are in pure AdS the allowed configurations are consistent with the system having a bare UV quark mass that scales as the inverse of the domain walls separation. 

We have then attempted to trigger chiral symmetry breaking in the system. An applied baryon number magnetic field, via the D7 world volume gauge field, turns out to not trigger chiral symmetry breaking. Inspired by the effective unbackreacted dilaton prefactor that the B field generates in the probe action we have found bottom up profiles that do cause condensation. We exhibited the IR mass gap formation and showed the Goldstone boson/pion of the system satisfies a Gell-Mann-Oakes-Renner relation. We have also studied the domain wall system in a background geometry with a backreacted dilaton flow. Whilst the geometry has a singularity which can not be clearly lifted in string theory it does provide a backreacted hard wall geometry that triggers chiral symmetry breaking on the domain walls. 

This paper has been a first exploration of these ideas. We have concentrated on the case of a single probe brane but if multiple coincident branes are used then the U(1) symmetries will be raised to U($N_f)$ chiral symmetries. If the flavours are degenerate then the scalar flavour multiplets masses just follow from the U(1) case. In the future we hope to construct a holographic domain wall model of QCD which would have these explicit non-abelian chiral symmetries. Beyond that we hope to find holographic duals of more complex chiral theories that can not currently be constructed. The domain wall trick seems like it will be a useful holographic tool for the construction of gauge theories.

\noindent {\bf Acknowledgements:} we are grateful for discussions with James Harrison.
JCR's work was supported by Mexico's National Council of Science and Technology (CONACyT) grant 439332, NEs  by the STFC consolidated grants ST/P000711/1 and ST/T000775/1. JMs work was supported by an STFC studentship.


\end{document}